\documentclass[10pt,a4paper,notitlepage,aps,pra,showpacs,superscriptaddress,floatfix,nofootinbib,twocolumn,accepted=2022-01-09]{quantumarticle}
\pdfoutput=1

\usepackage{amsmath,amsfonts,amsthm,amssymb}
\usepackage{graphicx}
\usepackage{xspace}

\usepackage{bbm}
\usepackage{color}
\usepackage{float}
\usepackage{algorithm}
\usepackage{algpseudocode}
\usepackage{diagbox}
\usepackage{appendix}

\usepackage[dvipsnames]{xcolor}
\definecolor{quantumviolet}{HTML}{53257F}
\usepackage[colorlinks=true,citecolor=Emerald,linkcolor=quantumviolet,urlcolor=quantumviolet,hyperindex]{hyperref}

\newtheorem*{theorem*}{Theorem}

\theoremstyle{definition}
\newtheorem*{definition}{Definition}
\theoremstyle{obs}
\newtheorem{obs}{Observation}

\theoremstyle{conjecture}
\newtheorem{conjecture}{Conjecture}

\newcommand{\id}{\mathbbm{1}} %identity operator

\newcommand{\T}{\mathsf T}

\newcommand{\va}{\vec{a}}

\newcommand{\etal}{\emph{et al.}\xspace}

\newcommand{\mean}[1]{\langle #1 \rangle}
\newcommand{\ketbrac}[1]{|#1\rangle\langle #1|} %ketbra

\newcommand{\aot}{{\va_{\rm ot}}}

\newcommand{\II}{\mathcal I}

\newcommand{\e}{e}

\DeclareMathOperator{\tr}{tr}
\DeclareMathOperator{\DC}{DC}
\DeclareMathOperator{\PC}{PC}

\newcommand{\floor}[1]{\left \lfloor #1 \right \rfloor}

\newcommand{\ket}[1]{| #1 \rangle}
\newcommand{\bra}[1]{\langle #1 |}
\newcommand{\ketbra}[1]{| #1 \rangle\langle #1 |}

\newcommand{\Ostar}{\Omega^{\star}(\va,d)}
\newcommand{\Oest}{\Omega^\text{est.}(\va,d)}

\newcommand{\appref}[1]{\hyperref[#1]{Appendix \ref*{#1}}}

\begin{document}
	\title{Temporal correlations in the simplest measurement sequences}
	
	\author{Lucas B. Vieira}
	\email{lucas.vieira@oeaw.ac.at}
	\affiliation{Institute for Quantum Optics and Quantum Information (IQOQI), Austrian Academy of Sciences,\\ Boltzmanngasse 3, 1090 Vienna, Austria}
	\orcid{0000-0002-6530-8271}
	\author{Costantino Budroni}
	\email{costantino.budroni@univie.ac.at}
	\affiliation{Faculty of Physics, University of Vienna, Boltzmanngasse 5, 1090 Vienna, Austria}
	\affiliation{Institute for Quantum Optics and Quantum Information (IQOQI), Austrian Academy of Sciences,\\ Boltzmanngasse 3, 1090 Vienna, Austria}
	\orcid{0000-0002-6562-7862}
	
	\begin{abstract}
		We investigate temporal correlations in the simplest measurement scenario, i.e., that of a physical system on which the same measurement is performed at different times, producing a sequence of dichotomic outcomes. The resource for generating such sequences is the internal dimension, or \emph{memory}, of the system. We characterize the minimum memory requirements for sequences to be obtained deterministically, and numerically investigate the probabilistic behavior below this memory threshold, in both classical and quantum scenarios. In the classical case, a particular class of sequences is found to offer an upper-bound for all other sequences, which  suggests a nontrivial universal upper-bound of $1/\e$ for the probability of realization of any sequence below this memory threshold. We further present evidence that no such nontrivial bound exists in the quantum case.
	\end{abstract}
	
	\maketitle
	
	\section{Introduction}
	Temporal correlations arise in any information processing task that consists of sequential operations, possibly involving inputs, producing an output at each time step. From a quantum mechanical perspective, these operations can be interpreted as measurements of some observables of a physical system. Models of quantum computation involve this notion ~\cite{MarkiewiczPRA2014,Zurel:2020PRL}, but a variety of different approaches are present in the literature: from random access codes (RACs) ~\cite{Wiesner1983,Ambainis1999, Ambainis2002, BowlesPRA2015, AguilarPRL2018, Miklin2020}, to classical simulations of quantum contextuality~\cite{KleinmannNJP2011,Fagundes2017}, quantum simulation of stochastic processes~\cite{GarnerNJP2017,Elliott2018,Elliott2019}, purity certification~\cite{SpeePRA2019}, and time-keeping devices~\cite{ErkerPRX2017,Woods2018,Woods2021,SchwarzhansPRX2021}.
	
	In parallel to quantum information applications, an analysis of nonclassical properties of temporal 
	correlations was proposed by Leggett and Garg (LG)~\cite{LeggettPRL1985,EmaryRPP2014}, who investigated the 
	limitation of temporal correlations in a classical theory they called \emph{macroscopic realism}. Similarly to 
	the case of Bell nonlocality~\cite{Bell1964,nonloc_rev}, an experimental violation of LG inequalities disproves 
	that the observed correlations come from a macrorealist theory. In addition to the standard assumption of 
	\emph{realism}, i.e., that physical variables have a definite value at any instant of time, macrorealist 
	theories assume that such values can be measured in a \emph{non-invasive} way. The latter is a strong 
	condition that makes LG experiments challenging due to the \emph{clumsiness loophole}~\cite{WildeMizel2012}: A 
	clumsy measurement alone may be responsible for the violation of the LG inequality, and considerable effort is 
	needed to close it~\cite{BudroniPRL2015,HalliwellPRA2016,KneeNC2016,EmaryPRA2017,UolaPRA2019}. It is clear, 
	then, that any information processing task where the information carrier is modified by sequential operations 
	cannot be discussed in terms of LG inequalities. A recently introduced framework~\cite{BudroniNJP2019} for 
	temporal correlations, admitting invasive measurements, overcomes this problem: any operation on the physical 
	system under consideration modifies (or {\it store information} in) it, but up to a limit, given by its 
	\emph{internal memory}. The case of a noninvasive measurement is recovered for zero bits of internal memory.
	Of course, the amount of invasiveness must be limited, otherwise classical theory is able to generate the 
	same correlations as quantum theory~\cite{FritzNJP2010,Hoffmann2018}. 	
	This framework has been explored from several perspectives:  quantum dimension 
	witnesses~\cite{Hoffmann2018,SpeeNJP2020b}, differences between classical, quantum, and general probability 
	theories ~\cite{BudroniNJP2019}, minimal dimensional realizations of extreme correlations~\cite{SpeeNJP2020} and 
	their convex mixtures~\cite{Mao2020} (albeit with different assumptions on the time evolution and convexity 
	properties), and time-keeping devices~\cite{Budroni2020}.
	
	In this paper, we consider the simplest form of temporal correlations: A single finite-dimensional system subjected to the same dichotomous measurement at different times. Such a system can be modeled as a probabilistic \emph{finite-state automaton}~\cite{rabin1963probabilistic,PazBook} (FA, defined in \autoref{sec:prel_not}).	First, we investigate the minimal dimension necessary to generate a sequence, which we call its \emph{deterministic complexity} ($\DC$), and provide an efficient algorithm to compute it. Below this dimension, i.e., for $d<\DC$, any realization must be probabilistic and only here difference between classical and quantum correlations may arise~\cite{SpeeNJP2020}. Then, we investigate numerically all classical sequences up to $L=10$, and all $d< \DC$. Our results suggest that the probability of each sequence, for any $d< \DC$, can be upper bounded by the probability of a special sequence, called \emph{one-tick sequence}, of length $\DC$ and realized in the same dimension. If proven, this conjecture would imply an universal bound of $1/\e$ for the probability of realization of any sequence in $d<\DC$. Finally, we show how quantum models can outperform classical ones and provide an explicit model approaching the bound of probability $1$ in the limit of long sequences and high dimension.
	
	The paper is organized as follows. In Sec.~\ref{sec:prel_not}, we introduce the formalism and basic concepts related to finite-state automata. In Sec.~\ref{sec:detseqs}, we investigate the minimal dimension necessary to deterministically generate a sequence of outputs. In Sec.~\ref{sec:probseqs}, we investigate minimal classical probabilistic realizations. In Sec.~\ref{sec:quantummodel}, we investigate quantum realizations. Finally, in Sec.~\ref{sec:conclusions}, we present the conclusion and outlook of our work.
	
	\section{Preliminary notions}\label{sec:prel_not}
	
	A finite-state automaton (FA) is a machine that generates an output, belonging to some alphabet $\mathcal{A}$, and performs a state transition both according to a time-independent probabilistic rule. We assume the machine has $d$ internal states (classical or quantum) and produces a sequence of outcomes $\va = (a_1, a_2, \dots, a_L)$,  $a_i \in \mathcal{A}$, for a sequence of measurements of length $L$. It may be helpful sometimes to denote the sequence of outcomes as $\va^{L}$ to indicate that it has length $L$. 
	
\begin{figure}[]\centering
		\includegraphics[width=1\linewidth]{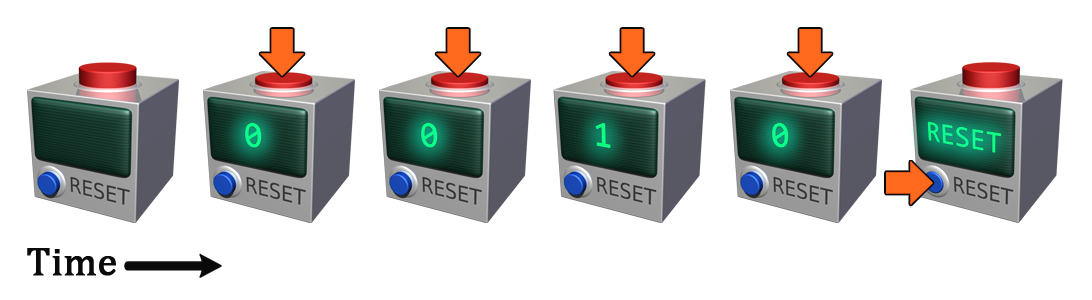}
		\caption{A pictorial representation of our physical model, as a box with a ``measure'' button (red, on top), a ``reset'' button (blue, in front), and a display for the outcomes of measurements performed by the box. The ``measure'' button is pressed $L$ times to generate a sequence of length $L$, then the ``reset'' button brings the machine to its initial state, and the measurement sequence can be repeated to collect statistics.}
		\label{fig:timebox}
	\end{figure}

	Our model may be thought of as a box with $d$ internal states, a display, a ``measure'' button, and a ``reset'' button (\autoref{fig:timebox}). Upon pressing ``measure'', the box performs a time-invariant, and possibly invasive, dichotomous measurement, and displays the outcome ($0$~or~$1$). We perform the measurement as many times as we wish, e.g., $L = 4$ times, obtaining a \emph{sequence} of outcomes, e.g., $\va = (0,0,1,0)$, after which we reset the box to its initial state and repeat the experiment. Over many trials we estimate a probability distribution over all possible outcomes the box can realize using its $d$ internal states as a resource.
	
	The classical model (see \cite{PazBook} for a textbook introduction, and \cite{Shallit2001automatic}) is described by a pair of matrices $T = (T_0, T_1)$, where $T_0,T_1$ are sub-stochastic transition matrices such that $T_0 + T_1$ is row-stochastic, i.e., with nonnegative entries, $[T_a]_{ij} \geq 0$ for $a=0,1$, and $\sum_j [T_0]_{ij} + [T_1]_{ij} = 1$ for all $i$. 
	Given a $d$-dimensional classical model $T$, a sequence $\va$, an initial state $\pi$, its probability can be computed as
	\begin{equation}
	p(\va|T,d) = \pi T_{a_1} T_{a_2} \dots T_{a_L} \eta,
	\end{equation}
	where $\pi$ is a distribution over all possible $d$ states, i.e., $\pi_i \geq 0$ and $\sum_i \pi_i=1$, and $\eta:=(1,1,\ldots,1)^\T$ provides the sum over all possible final states. From this definition, it is clear that switching all $0 \leftrightarrow 1$ gives the same probability, as the model is symmetric under relabeling. Thus, without loss of generality, we can assume $a_1 = 0$.
	
	In the quantum case, the output generation is simply interpreted as sequential quantum measurements. The output probability and state transition are described by the instruments $\II = (\II_0, \II_1)$, in the Heisenberg picture, i.e., $\II_a$ is completely positive (CP) for $a=0,1$ and $\II_0 + \II_1$ is a unital map. For an initial state $\rho$ on a $d$-dimensional Hilbert space, the probability for a sequence  $\va$ is then
	\begin{equation}
	p(\va|\II,d) = \tr[\rho\ \II_{a_1} \circ \II_{a_2}\circ \ldots \II_{a_L} (\openone) ].
	\end{equation}
	The classical case is recovered by requiring the initial state and transformation to be diagonal in the same basis.
	The time evolution is not explicitly considered here. However, a unitary time evolution, and even some forms of Markovian ones, can be absorbed into the definition of the instruments with a proper choice of measurement times, and deviations from this ideal situation can be taken into account~\cite{SpeeNJP2020}. 
	
	We call the pair $T=(T_0,T_1)$ or the pair $\II = (\II_0, \II_1)$, a FA, or simply a \emph{model}, when its classical or quantum nature is clear from the context or irrelevant.
	
	A sequence of particular interest, as we shall see, is the sequence $\va^L=(0,0,\ldots,0,1)$. It appeared in the investigation of classical and quantum clocks~\cite{Budroni2020}, as the sequence corresponding to one tick of the clock. We, thus, call it the \emph{one-tick sequence} and denote it as $\aot^L$.
	
	\section{Lossless compression of a model}\label{sec:detseqs}
	
	\begin{figure}[t]\centering
		\includegraphics[width=1\linewidth]{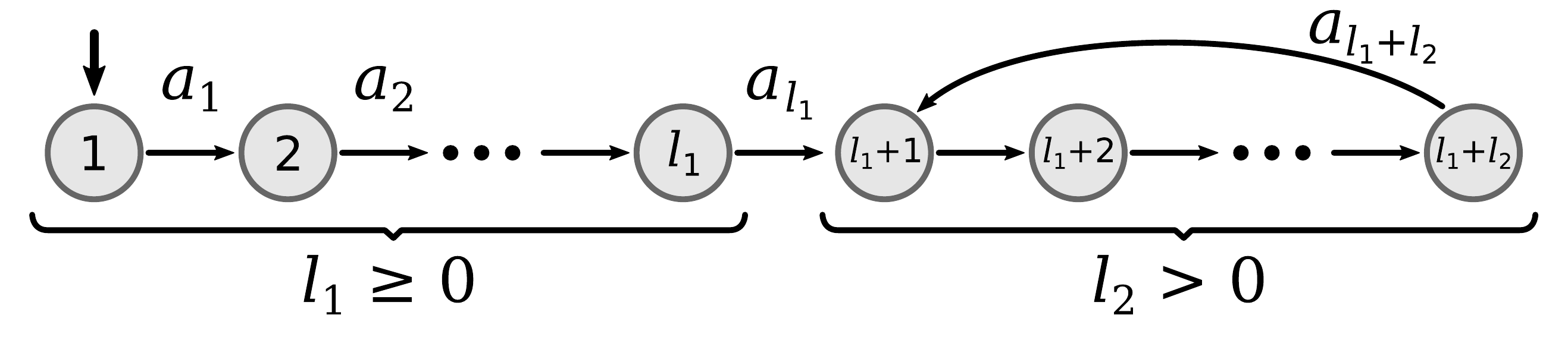}
		\caption{The general structure of minimal DFAs corresponding to minimal patterns.  The initial state is denoted by a thick incoming arrow, nodes of the graph represent the internal states, arrows state transitions and symbols $a_i$ the output associated with them. For a pattern the minimal model consists of a tail, with $l_1 \ge 0$ states, which leads to a cycle, with $l_2 > 0$ states.}
		\label{fig:dfa_diagram}
	\end{figure}
	
	The internal memory of the system, either classical or quantum, is the resource needed to reproduce a given sequence $\va$. Not all models, however, use this memory efficiently. In many cases, the number of internal states of a model can be reduced, still obtaining the chosen sequence with probability $1$. In this section, we investigate precisely the minimal amount of resources needed to reproduce a sequence deterministically, by providing an explicit algorithm to compute it. We first outline our working definition.
	
	\begin{definition} (Lossless compression of a model). If the number of internal states of a model can be reduced, and the model is still able to generate the same sequence with probability $1$, we speak of a \emph{lossless compression} of the model, where we use the term ``compression'' strictly in the sense of number of states in a model.
	\end{definition}
	
	This information is central to the characterization of the difference between classical and quantum correlations in time, as quantum deterministic models require no coherence, hence, they can be simulated by classical models of the same dimension. In fact, as shown in \cite{SpeeNJP2020}, states that produce a different sequence of outcomes with probability one are orthogonal, hence, a minimal representation can be realized with pure states and deterministic transitions among them. This quantum model, then, requires no coherence and can be simulated classically. For this reason, in the following we simply discuss classical models in the investigation of the minimal dimension necessary for a deterministic realization of a sequence.
	
	We remark that this is in stark contrast with other models of temporal correlations, such as the original Leggett-Garg formulation~\cite{LeggettPRL1985,EmaryRPP2014}, which assumes noninvasive measurability, quantum contextuality for sequential measurements~\cite{Kirchmair:2009NAT,Guhne:2010PRA,Context_review}, which assumes some form of compatible measurements, or other forms of temporal quantum correlations assuming projective measurements~\cite{BudroniPRL2013,BudroniPRL2014,SchildPRA2015,RingbauerQ2017,Sohbi2021} or a combination of spatial and temporal correlations~\cite{GallegoNJP2014,Spee2020arxiv,BowlesQ2020}.
	
	Given these considerations, we introduce the following notion.
	
	\begin{definition}\label{def:dc}
		(Deterministic Complexity). Given a sequence $\va$, we define its \emph{deterministic complexity}, denoted by $\DC(\va)$ or simply DC, as the minimal $d$ such that there exists a model $T$ giving $p(\va|T,d)=1$.
	\end{definition}
	
	A trivial upper bound for $\DC$ is given by the sequence length, i.e., $\DC(\va^L)\leq L$. The bound is generally not tight: it is saturated only in the case of the one-tick sequence, i.e., $\DC(\aot^L) = L$ (see \appref{app:dcpatterns}) and the equivalent one obtained by the substitution $0\leftrightarrow 1$.
	A general criterion for counting the minimal number of states necessary to realize a given sequence deterministically has been formulated in Ref.~\cite{SpeeNJP2020}. In simple terms, one may say that two different steps of a sequence are associated with the same internal state if ``they have the same future'', namely, if they are followed by the same sequence of outcomes. The minimal number of internal states is the number of different, or inequivalent, futures. This also give us the intuition as to why the one-tick sequence is the one with the highest deterministic complexity: each step $k<L$ is associated with the different future ``emit $1$ in $L-k$ steps'' (see also \appref{app:dcpatterns}). In the following, we present an efficient algorithm to compute such a minimal number of states for a FA.
	
	We have seen that each deterministic sequence can be realized by transitions among orthogonal states and that, for a minimal realization, rank-1 states are sufficient (see Ref.~\cite{SpeeNJP2020}). A simple argument shows that all minimal models are of this form, namely, that the evolution is always between rank-1 states and never creates a mixed state. Since, we are discussing transitions among orthogonal blocks, it is sufficient to consider the classical model.
	
	% ---- Observation
	\begin{obs}\label{obs:dettrans} A deterministic probability for a sequence $\va$ implies deterministic state transitions for its minimal (classical) model $T$, with $d = \DC(\va)$, i.e.,
		\begin{equation}\label{eq:dettrans}
		p(\va|T,\DC(\va)) = 1 \implies [T_a]_{ij} \in \{0,1\}, \; \forall\, i,j,a.
		\end{equation}
		\begin{proof}
			By contradiction, suppose that at the $n$-th transition, $n<L$, instead of moving deterministically, the automaton transitions to the state $s$ with probability $q$ or to the state $s'$ with probability $(1-q)$. The total probability, then, can be written as $1=p= q p_1 + (1-q) p_2$, where $p_1$ and $p_2$ are the probabilities for the two paths, conditioned on that probabilistic transition at the $n$-th step. Since $0\leq q \leq 1$ and $p_i\leq 1$, $p=1$ implies $p_1=p_2=1$. Hence, it is sufficient to follow the path going through, e.g., $s$ to generate the sequence with probability $1$. One may, then, simply remove the state $s'$ and put $q=1$, in contradiction with the assumption that the dimension was minimal. The argument can be applied iteratively, if more than one probabilistic transition appears.
			The only remaining case is that of a transition happening in the last step, i.e., $n=L$, in which case the transition is irrelevant for the model. In the terminology above,  $1=p=p_1(q+1-q)$. 
		\end{proof}
	\end{obs}
	
	% ---------- UNTIL HERE
	Notice that the previous argument uses only the probability over paths in the state space, thus, it is valid also for the case of machines accepting inputs as in, e.g., Refs.~\cite{Hoffmann2018,BudroniNJP2019,SpeeNJP2020}.
	
	Such deterministic models are referred to as {\it deterministic finite-state automata} (DFA). These DFAs 
	have a special form that allows for an easier characterization. In such cases, the automata are able to generate 
	with probability one not only the finite sequences considered, i.e., $p(\va|T,\DC(\va))=1$, but an infinite 
	family of sequences. For instance, the sequence $\va = (0,0,0,1,1,0,1,1,0)$ can be seen as a truncation of the 
	infinite sequence $\mathtt{00(011)}$, where $\mathtt{(011)}$ denotes that the subsequence $\mathtt{011}$ is 
	repeated indefinitely. We call such a description of infinite sequences a \emph{pattern}, and to each pattern 
	corresponds a deterministic model. In this simple example, the pattern consists of an initial sequence which 
	occurs once, the \emph{tail} $\mathtt{00}$, and the \emph{cycle} $\mathtt{(011)}$, which occurs at least once, 
	while further repetitions may be truncated. Since the state transitions are deterministic, it is clear that to 
	each output is associated the current state of the machine and the transition to the subsequent one.  This structure is general: Since the number of states $d$ is finite for $L \geq d$, the system must 
	at some point transition back to a previously used state and repeat the sequence thereafter, as in 
	\autoref{fig:dfa_diagram}. This can be summarized in the following
	
	\begin{obs}
		Every minimal deterministic model is characterized by a {\it tail} and a {\it cycle}, which completely describe the structure of the state transitions and generate a {\it pattern}. The minimal number of states needed to generate a pattern is precisely the length of the pattern. 
	\end{obs}
	Note that, due to truncations of the cyclic part, more than one minimal pattern may describe a given sequence, but all such patterns share the same length. Clearly, a DFA generating a minimal pattern of length $\ell$ is minimal if it uses exactly $\ell$ states.
	
	This suggests an algorithm to compute $\DC$ for an arbitrary sequence $\va^L$: it is enough to compute the length of a minimal pattern that generates it. Intuitively, for a sequence $\va^L$ its minimal patterns are characterized by two numbers $l_1$ and $l_2$, respectively, the length of the tail and the length of the cycle, such that $\DC(\va) = l_1 + l_2$. An explicit algorithm to compute minimal patterns is presented in \appref{app:dcpatterns}. Here, we present a brief outline.
	
	\smallskip
	\noindent \textbf{Algorithm (outline)}. To find the minimal patterns and the deterministic complexity of a given sequence $\va$ of length $L$:
	\begin{enumerate}
		\item Assume a pattern length $\ell = 1,\dots,L$. To ensure minimality of the pattern, we test each $\ell$ in increasing order.
		\item For each $\ell$, assume a tail length $l_1 = 0,\dots,\ell-1$, giving a cycle length $l_2 = \ell - l_1$.
		\item If all outcomes $a_i$ for $i = \ell+1,\dots,L$ can be interpreted as repetitions of the cycle subsequence, $(a_{l_1 + 1}, \dots, a_\ell)$,  then $(l_1,l_2)$ describes a valid pattern for $\va$ of length $\ell$, and thus, $\DC(\va) = \ell$.
		%  $a_{\ell+i} = a_{l_1 + 1 + (i \mod l_2)}$ for all $i=1,\dots,L-\ell$, then $(l_1,l_2)$
		\item If we also wish to find all patterns compatible with $\va$, we may continue checking the remaining $l_1$ for the same $\ell$.
	\end{enumerate}
	In this way, we need to generate and compare at most $L^2$ patterns with the original sequence, corresponding to $O(L^3)$ operations.
	
	Interestingly, the tail and cycle structure also allows us to compute the exact number of minimal patterns that can be generated with exactly $\ell$ states, adapting an argument by Nicaud~\cite{Nicaud1999} for unary automata, based on two simple conditions:  $(i)$ the tail is minimal, and $(ii)$ the cycle is minimal. The only nontrivial counting corresponds to that of minimal cycles, which can be mapped to a known combinatoric problem of \emph{primitive words}, i.e., strings of symbols $w$ that cannot be written as $w=x^k$ for some $k\geq 2$~\cite{allouche2003automatic}. An explicit expression for the number of minimal patterns length $\ell$ over a $k$-symbol alphabet is given by
	\begin{equation}\label{eq:npatt}
	N_k(\ell) = \psi_k(\ell) + \sum_{i=1}^{\ell-1} (k-1) k^{i-1} \psi_k(\ell-i),
	\end{equation}
	where $\psi_k(n) = \sum_{d | n} \mu(d) k^{n/d}$ is the number of primitive words of length $n$ in a $k$-symbol alphabet and $\mu(d)$ is the M\"obius function. The details of the derivation are presented in \appref{app:dcpatterns}. 
	
	In summary, the deterministic complexity of a sequence represents the optimal ``lossless compression'' of its finite-state automaton. It provides the threshold after which quantum systems provide an advantage: namely for ``lossy compression'' of a sequence $\va$, corresponding to $d<\DC(\va)$. It is also connected to what appears to be a classical upper bound for this probability, which depends only the pair $(d, \DC(\va))$ and can be computed in terms of the one-tick sequence $\aot^{\DC(\va)}$. This potential upper bound is discussed in the next section.
	
	Finally, it is worth commenting on the terminology used. The number of internal states of an automaton is commonly used as a measure of complexity, such as \emph{state complexity} for a regular language~\cite{Holzer_Review, Nicaud1999}, the \emph{statistical complexity} associated with an $\varepsilon$-machine~\cite{ShaliziJSP2001}, or \emph{automatic complexity}~\cite{Shallit2001automatic}. By \emph{compression} of a model, we simply mean the reduction of its number of internal states. Of course, this ``compression'' cannot be interpreted as a compression of the bit-string of outputs itself, as the description of the model may simply be more expensive, in terms of memory, than the original sequence. Since the stopping condition (i.e. the length $L$ of the sequence) is specified \emph{a priori}, and not embedded in the automaton, the deterministic complexity as defined herein possesses different properties with respect to the other notions in the literature.
	
	\section{Lossy compression of a classical model}\label{sec:probseqs}
	
	\begin{figure}[t]\centering
		\includegraphics[width=0.9\linewidth]{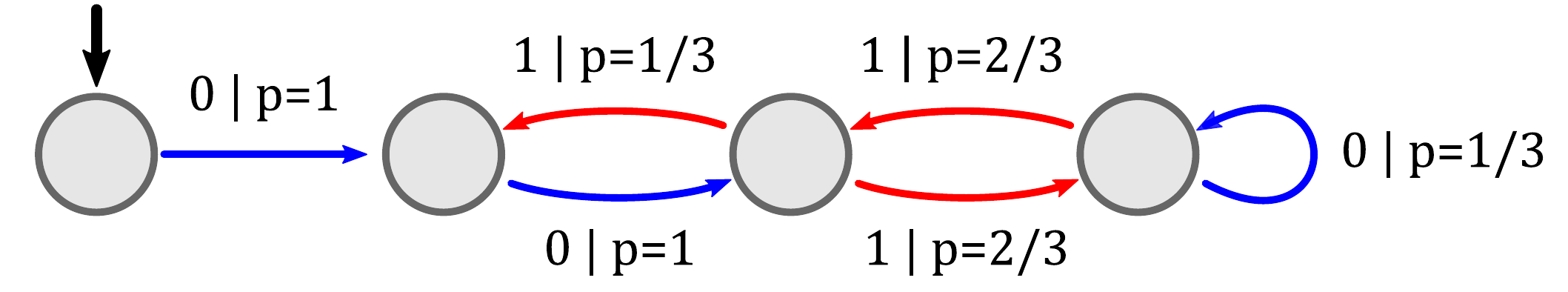}
		\caption{The optimal model found for the sequence $\va = (0,0,1,0,1,1)$ ($\DC(\va) = 5$) and $d = 4$, which results in $p = 8/27 = 0.\overline{296}$. Arrow colors were used to help distinguish the $0$ (blue) and $1$ (red) transitions, and labels correspond to their respective outputs and transition probabilities, e.g., ``$1 \vert p=2/3$'' denotes an output of \texttt{1} with probability $2/3$. The model contains a nontrivial combination of probabilistic and deterministic transitions. Different sequences can lead to vastly different models, and the structure of such behaviors are still under study.}
		\label{fig:pa_model}
	\end{figure}
	
	In this section, we investigate the optimal probability with which a sequence $\va$ can be generated when $d< \DC(\va)$. In line with the previous discussion, this procedure can be interpreted as a further compression of the model, beyond the threshold $d=\DC(\va)$. By definition, we necessarily have $p < 1$ in this case, and therefore we refer to this as a \emph{lossy compression} of a model.
	
	\subsection{Survey of general sequences}
	
	The optimal models obtained under such constraints must make nontrivial use of the memory resources available, which gives rise to complex behaviors and transitions that are specific for each sequence, as shown in \autoref{fig:pa_model} and \autoref{fig:neardc}.
	
	In studying these behaviors, we looked at all $2046$ sequences $\va^L$, for $L=2,3,\ldots,10$, and all possible realizations in dimensions $d$ such that $d<\DC(\va^L)$. To do so, we analyzed the nonconvex constrained problem
	\begin{equation}\label{eq:optprob}
	\begin{split}
	\max_{\{T\} }\ & p(\va|T,d) = \pi_0 T_{a_1} T_{a_2} \dots T_{a_L} \eta,
	\\
	\text{subjected to: }& \sum_{j} \left( [T_0]_{ij} + [T_1]_{ij} \right) = 1, \forall\, i,
	\\ &  [T_a]_{ij} \ge 0,\ \forall i,j,
	\end{split}
	\end{equation}
	via gradient-descent methods, in particular, the algorithm Adam~\cite{Adam}. This is possible by transforming the problem into an unconstrained one, see \appref{app:survey} for details.
	
	The results of this general survey highlighted a special property of the one-tick sequences, which required further investigation. This is described next.
	
	\subsection{Survey of one-tick sequences}
	
	In addition to this extensive search, the one-tick sequences in particular were also surveyed based on an improvement of the optimal classical models extensively investigated in Ref.~\cite{Budroni2020}, both with analytical and numerical methods. There, a specific model referred to as the \emph{multicyclic model} was shown  to be optimal in some cases and thought to be optimal in general.
	Here, we show that this model can be improved with a slight modification, and in several cases such models increase the probabilities previously obtained. We term these \emph{enhanced multicyclic models} (EMCMs), in analogy to the multicyclic model in Ref.~\cite{Budroni2020}. However, the optimality of these models is still conjectured.
	
	\begin{figure}[h]\centering
		\includegraphics[width=\linewidth]{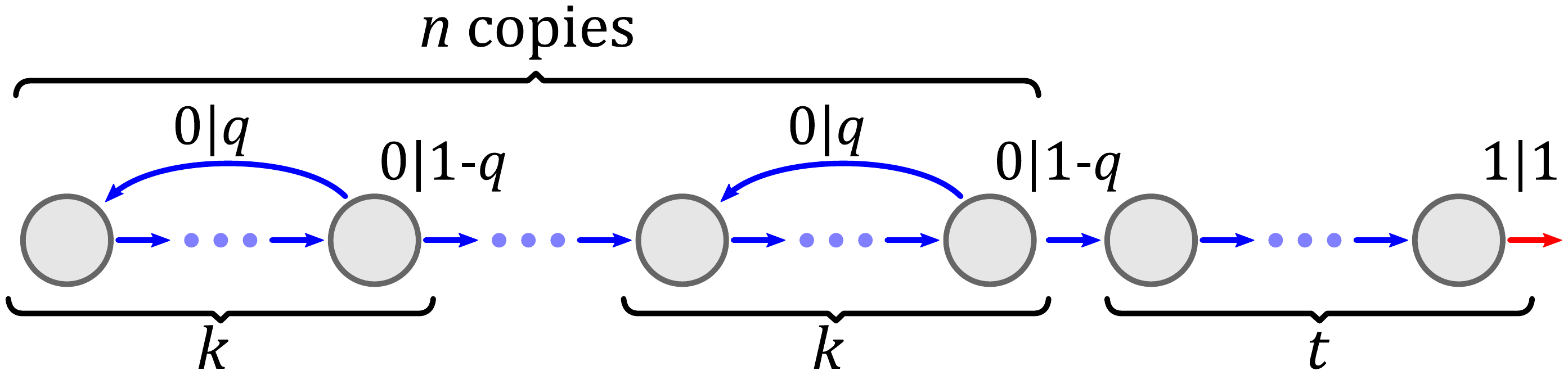}
		\caption{The structure of enhanced multicyclic models, consisting of $n$ identical probabilistic cycle blocks of size $k \in \mathbb{N}$, followed by $t \ge 0$ deterministic transitions such that $d = n k + t$. The transitions within each cycle are deterministic. The initial state (not shown) can be any of the first $k$ states in the first block.}
		\label{fig:emcm}
	\end{figure}
	
	A schematic representation of EMCMs is presented in \autoref{fig:emcm}. The model consists of $n$ blocks of size $k$, each forming its own cycle, followed by $t \ge 0$ deterministic transitions, such that $d = nk + t$. The transitions within each cycle block are deterministic. In the last state of each cycle, the machine can either cycle through the block with probability $q$, or transition to the next one with probability $1-q$. All these transition are associated with the output $0$, and once the last state of the last cycle is reached, if $t > 0$, the machine can either cycle again with probability $q$, or step into the final block with deterministic transitions. At the end of the final $t$ deterministic transitions, the machine emits the outcome $1$ with probability $1$, after which the subsequent state transition is irrelevant. In the special case where $t = 0$, and there are no deterministic transitions at the end, the output $1$ is associated to the forward transition of the last cycle.
	
	As an example, an EMCM with $k=2, n=2, t=1$ is as following:
	\begin{equation}\label{eq:TEMCM}
	T_0 = \left[
	\begin{array}{cc|cc|c}
	0 & 1 & 0 & 0 & 0 \\
	q & 0 & 1-q & 0 & 0 \\ \hline
	0 & 0 & 0 & 1 & 0 \\
	0 & 0 & q & 0 & 1-q \\ \hline
	0 & 0 & 0 & 0 & 0 \\
	\end{array}
	\right]\ , \ 
	T_1\eta = \left[
	\begin{array}{c}
	0\\
	0\\
	0\\
	0\\
	1
	\end{array}
	\right] \ . 
	\end{equation}
	
	As initial state, one can choose any state within the first block, otherwise the states of previous blocks are never used, in contradiction with the assumption of a minimal model. The optimal probability $q$ depends on the structure of the model and the length of the sequence.
	
	The special case $n=d, k=1, t=0$ corresponds to the {\it one-way} model $T_{\rm ow}$ described in \cite{Budroni2020}, for which the optimal probability $p(\aot^L|T_{\rm ow}, d)$, for $d<L$, is given by the negative binomial distribution:
	\begin{equation}\label{eq:fowmain}
	F_{\rm ow}(L,d)=\binom{L-1}{d-1} \left(1-\frac{d}{L}\right)^{L-d}\left(\frac{d}{L}\right)^d,
	\end{equation}
	where the optimal cycle probability is $q = 1 - d/L$.
	An EMCM can be described by just five parameters
	\begin{equation}\label{eq:emcm_params}
	\begin{aligned}
	&L &\text{the sequence length for the model},\\
	&n &\text{the number of cycle blocks},\\
	&k &\text{size of the cycle blocks},\\
	&t &\text{size of the deterministic block},\\
	&z &\text{initial state shift}
	\end{aligned}
	\end{equation}
	with $L,n,k,t,z\in \mathbb{N}$, $k>0$, $n = (d-t)/k$, and $z \leq k-1$, where $z=0$ means we start from the initial state, $z=1$ from the second, and so on. Note that only $k$ and $z$ are independent parameters of the model, whereas $n$ and $t$ depend jointly on the model and the sequence. We take Eq.~\eqref{eq:emcm_params} as a definition of the EMCM.
	
	Denoting the corresponding model as $T^{(L,n,k,t,z)}$, we can compute the probability for the $\aot$ sequence as
%	\begin{equation}\label{eq:opt_ow}
%	p(\aot^L|T^{(L,n,k,t,z)}, d) = F_{\rm ow}\left( \frac{L-t+z}{k},\frac{d-t}{k} \right),
%	\end{equation}
	\begin{align}\label{eq:opt_ow}
		p(\aot^L|T^{(L,n,k,t,z)}, d) = F_{\rm ow}\left( L' , d' \right) \\
		\text{with}\quad L' = \frac{L-t+z}{k}, \quad d' = \frac{d-t}{k} \notag
	\end{align}
	where, in addition to previous constraints, $z$ satisfies $(L-t+z)/k\in\mathbb{N}$. Given Eq.~\eqref{eq:opt_ow}, we may thus optimize directly the parameters $k$ and $z$ and obtain $(L,n,k,t,z)$ for any $(L,d)$. More details and intuitions for these results, in particular the meaning of $(L',d')$, can be found in \appref{app:otseqs}. We will refer to optimal EMCMs, for a given context, simply by $E$. 
	
	\medskip
	
	%%%%%%%%%%%%%%%%%%%%%%%%%%%%%%%%%%%%%%%%%%%%%%%%%%%%%%%%%%%%%%
	
	\begin{figure}[h]\centering
		\includegraphics[width=1\linewidth]{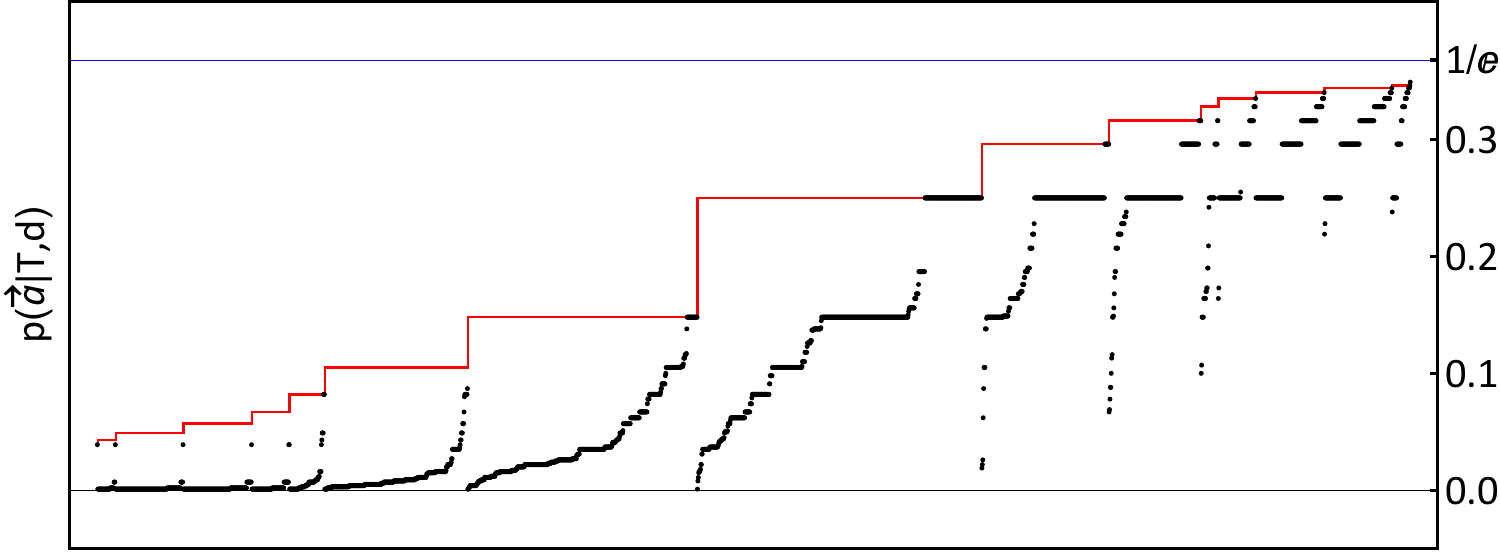}
		\caption{Probabilities of optimal classical models (black dots) for all $L = 10$ sequences and all $d < \DC(\va)$. The red line is the EMCM upper-bound as described by Conj.~\ref{conj:main}. For clarity, the points have been sorted by increasing conjectured bound, then by increasing probability. At the top, the universal classical bound $1/\e$ is shown.}
		\label{fig:emcm_upperbound_l10}
	\end{figure}
	
	%%%%%%%%%%%%%%%%%%%%%%%%%%%%%%%%%%%%%%%%%%%%%%%%%%%%%%%%%%%%%%
	
	\subsection{One-tick sequence probability as an upper-bound}
	
	An example of the numerical results of both surveys, for the case of $L=10$, is presented in \autoref{fig:emcm_upperbound_l10}. For general sequences and optimal EMCMs for one-tick sequences, the results have revealed a striking property of the one-tick sequences: 
	%they seem to achieve the highest probability out of any sequence if $d  < \DC(\va)$ and $L \le \DC(\va)$.
	%In particular, 
	The optimal probability for the one-tick sequence of length $\DC(\va)$ at dimension $d$ seems to act as an upper-bound for all sequences with the same $\DC$ and $d$.  While a proof of this result is yet to be found, our current results led us to formulate the following conjecture:
	
	\begin{conjecture}\label{conj:main}The optimal probability of any sequence $\va$, in any classical $d < \DC(\va)$ scenario, is upper-bounded by the probability of the one-tick sequence $\aot^{L}$ with length $L = \DC(\va)$, using the EMCM on $d$ states, i.e.:
		\begin{equation}\label{eq:conj}
		p(\va^{L}|T,d) \le p(\aot^{\DC(\va^{L})}|E,d), \quad \forall \, d < \DC(\va^L)
		\end{equation}
		where $E$ is the optimal EMCM model with $d$ states.
	\end{conjecture}
	
	In light of this, and in analogy with $\DC$, we may also define the $q$-\emph{probabilistic complexity} of a sequence $\va$ for $q\in[0,1]$, denoted as $\PC_q(\va)$, as the minimal $d$ such that there exists a model $T$ giving $p(\va|T,d) \ge q$. By definition, $\PC_1 = \DC$. The previous conjecture, then, would imply a stronger relation which is presented in the following.
	
	\begin{obs}\label{obs:c1c2} Provided Conj.~\eqref{conj:main} holds, then it follows that:
		\begin{itemize}
			\item $\PC_q(\va^L) = \DC(\va^L), \forall\ L, \va^L,$ if $q \geq 1/\e$, 
			\item[] or equivalently,
			\item $p(\va^{L}|T,d) < 1/e, \forall\ L,\va^L,T,$  if $d< \DC(\va^L)$.
		\end{itemize}
		\begin{proof}
			To show that it is sufficient to notice that 
			\begin{equation}
			\begin{split}
			p(\aot^L|E, d)\leq \sup_{L',d'<L'} F_{\rm ow} (L',d')
			= \sup_{L'} F_{\rm ow} (L', L'-1) \\ = \lim_{L'\rightarrow \infty} \left(1 - \frac{1}{L'}\right)^{L'}= \frac{1}{\e}\ ,
			\end{split}
			\end{equation}
			where we used Eq.~\eqref{eq:opt_ow} as a value for the EMCM, the fact that $F_{\rm ow} (L,d)$ is monotonically increasing in $d$ and the convergence of $ \left(1 - 1/L\right)^{L}$ is monotone, i.e., $1/\e$ is the upper bound for all $L$.
		\end{proof}
	\end{obs}
	
	In other words, any probabilistic realization with probability $q \geq 1/\e$ requires at least the same dimension as a deterministic realization, or equivalently, there is a universal upper bound probability for the classical realization of any sequence.
	
	These ideas are concisely shown in \autoref{fig:emcm_upperbound_l10} and \autoref{fig:emcm_upper_bounds}. Interestingly, for many different sequences and different optimal models the same few values appears as bounds, which are identical to those associated with the one-tick sequence. This is a hint of a rich structure that needs to be explored with more refined mathematical tools. For more details see \appref{app:survey}.
	
	The code used for the optimizations presented here and the data obtained as a result are publicly available at the online repository~\cite{githuburl}.
	
	%%%%%%%%%%%%%%%%%%%%%%%%%%%%%%%%%%%%%%%%%%%%%%%%%%%%%%%%%%%%%%
	
	\section{Quantum violations}\label{sec:quantummodel}
	
	%%%%%%%%%%%%%%%%%%%%%%%%%%%%%%%%%%%%%%%%%%%%%%%%%%%%%%%%%%%%%%	
	
	For the quantum models, one may describe the instruments $(\II_0,\II_1)$ by Kraus operators $K_a^i$ satisfying $\sum_{a,i} (K_a^i)^\dagger (K_a^i) = \id$. While the greater number of degrees of freedom and the non-linear constraints make a general optimization more difficult to perform and interpret, it is still instructive to investigate.
	
	\subsection{Survey of general sequences}

	For simplicity, we fix the dimension $d$ and the same number $N_K$ of Kraus operators for each instrument $\II_a$, such that
	\begin{equation}
		\II_a(\cdot) = \sum_{i=1}^{N_K} \left(K_a^i\right)^\dagger \cdot \left(K_a^i\right).
	\end{equation}
	
	For the numerical survey's quantum models, we looked at all $252$ sequences $\va^L$, for $L=2,\ldots,7$, and all possible realizations in dimensions $d$ such that $d < \DC(\va^L)$, and for $N_K = 1, 2, 3$. The constrained problem is then given by 
	\begin{equation}
		\begin{split}
			\max_{\II}\ & p(\va|\II,d) = \tr[\rho\ \II_{a_1} \circ \II_{a_2}\circ \ldots \II_{a_L} (\openone) ].
			\\
			\text{subjected to:} & \sum_{a,i} (K_a^i)^\dagger (K_a^i) = \id,
		\end{split}
	\end{equation}
	with $\rho = \ketbra{0}$, which we optimized via the Adam algorithm through an unconstrained form of the problem. See \appref{app:quantummodels} for details. Quantum advantages over the classical models were observed for all sequences, and the conjectured universal upper-bound of $1/e$ is violated by many sequences, but not all of them.
	
\begin{figure}[t]\centering
	\includegraphics[width=1\linewidth]{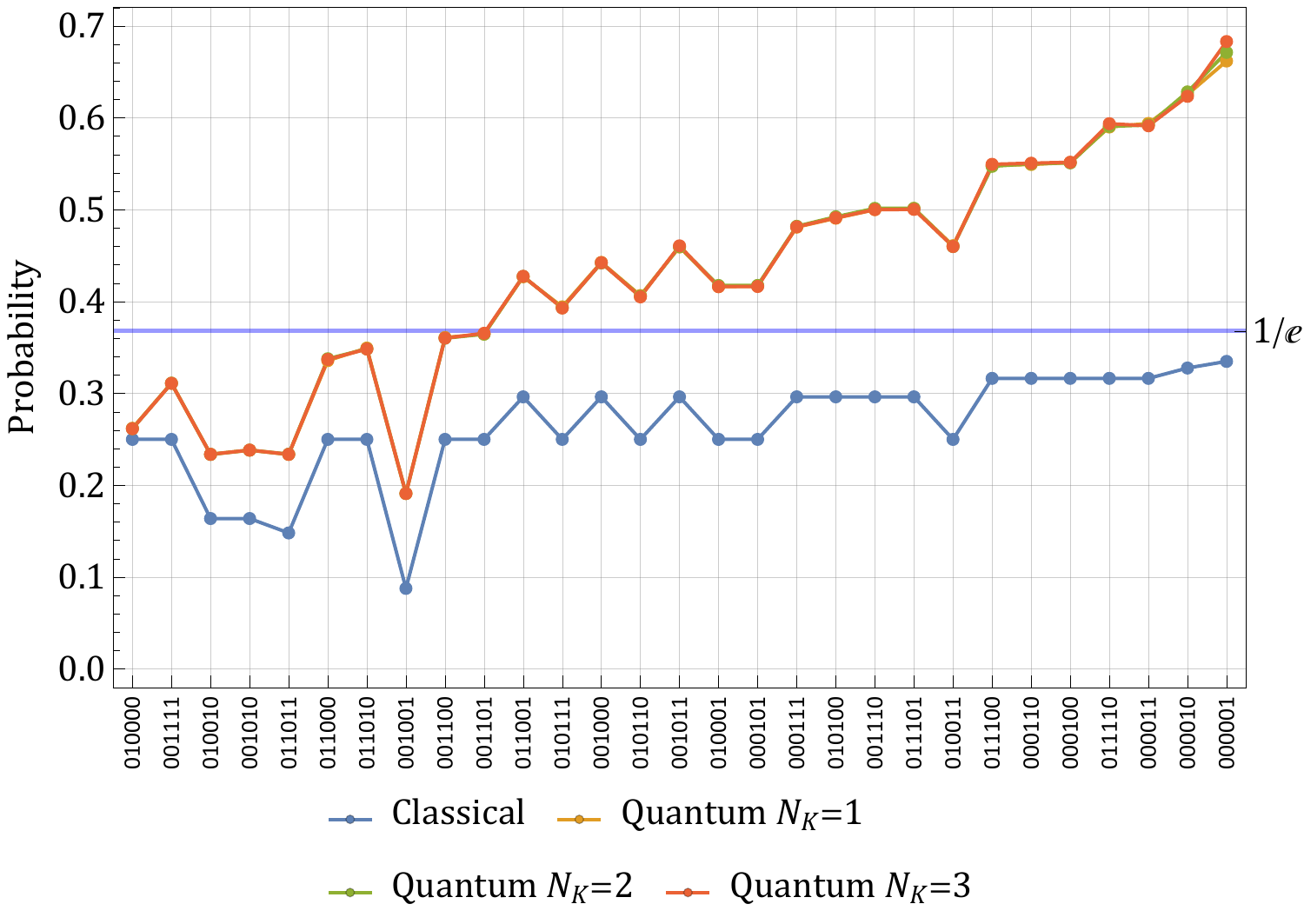}
	\caption{Probabilities found for $L = 6$, $d = \DC(\va) - 1$ using the quantum models described, sorted by increasing quantum advantage over the best classical model found. Note the coincidence of the values obtained for $N_K = 1, 2, 3$, up to numerical error, indicating no advantage in utilizing more than one Kraus operator. See \appref{app:quantummodels} for the full results.}
	\label{fig:quantumviol}
	\end{figure}

	A sample of our results is shown in \autoref{fig:quantumviol}. These results indicate that the one-tick sequence also outperforms all other sequences in the quantum case, while also providing the greatest advantage over its classical counterpart.
	Surprisingly, the results also display no benefit in using more than a single Kraus operator, as clearly seen in \autoref{fig:quantumviol} by the close coincidence of the results for each $N_K$. The optimization algorithm eventually converges with the various Kraus operators, for each instrument, being merely scalar multiples of one another.
	
	Naturally, classical models can be simulated by quantum ones, but classical probabilistic models always transform a pure state into a mixed state, what requires the use of multiple Kraus operators. With this in mind, a possible interpretation of this result is the observation that any quantum advantage would arise from the use of coherences between memory states, and these coherences are reduced as the state becomes mixed. Thus, these additional Kraus operators can only lower the performance of the model, by bringing their behavior closer towards the classical probabilistic behavior.
	
	\subsection{No nontrivial universal quantum bound}
		
	A natural question arises of whether a universal bound also appears in the quantum case, as in the classical scenario. To obtain some intuition on this problem, we analyzed the $\aot^L$ sequence for the special case $L=d+1$, i.e., ${d=\DC(\aot^L)-1}$. Despite the complexity of the problem, especially in high dimensions, this analysis is still feasible.
	
	For simplicity, we opt for using a single Kraus operator representation for each instrument, each with one degree of freedom according to the following parameterization. To automatically satisfy the constraints, the Kraus operators are written as the polar decomposition $K_a = U_a \sqrt{E_a}$, with $E_a \geq 0$ and $E_0 + E_1 = \id$. We can choose the operators $E_a$ to be diagonal and tailored for the target sequence (in our case, $\aot$), based on the optimal classical model found, i.e., $E_a = \operatorname{diag}(\eta_a)$, where $\eta_a = T_a \eta$ for the optimal $T_a$ of the classical model. As in \cite{Budroni2020}, the one-parameter unitaries are defined in terms of a Fourier transform of the computational basis, namely,
	\begin{equation}\label{eq:U_a}
		U_a = \e^{-\mathrm{i} H \theta_a},\qquad H = F \left( \sum_{i=0}^{d-1} k \ketbra{k} \right) F^\dagger.
	\end{equation}
		
	The only parameter of the model is then the angle $\theta_0$ appearing in Eq.~\eqref{eq:U_a}, since $E_0$ is fixed to be the diagonal matrix ${\rm diag}(1,1,\ldots,1,q)$, with $q$ taken as in the optimal classical model previously described, i.e., $q=1-d/(d+1)$.
	
	Let us formulate the problem in this simplified form. Fixing the dimension by the condition $d=L-1$ and an initial state $\rho=\ketbra{0}$, without loss of generality, the probability can be written as
	\begin{align}\label{eq:potq_m}
	p(\aot^L) &= \tr\left[ K_0^d \ketbra{0} (K_0^\dagger)^d E_1\right] = \tr\left[  \ketbrac{0} (K_0^\dagger)^d E_1 K_0^d \right] \notag \\
	\notag &= (1-q) \tr\left[  \ketbra{0} (K_0^\dagger)^d \ketbra{d-1} K_0^d \right]\\
	&= (1-q) \left| \mean{d-1| K_0^d |0}\right|^2,
	\end{align}
	where we used the cyclicity of the trace, and the definition of $E_1$ as $\id- E_0=(1-q)\ketbra{d-1}$. Using the optimal $E_0,E_1$, corresponding to $q=1-d/L=1-d/(d+1)$, we are left with only one parameter $\theta_0$ to  optimize.
	
\begin{figure}[]\centering
		\includegraphics[width=1\linewidth]{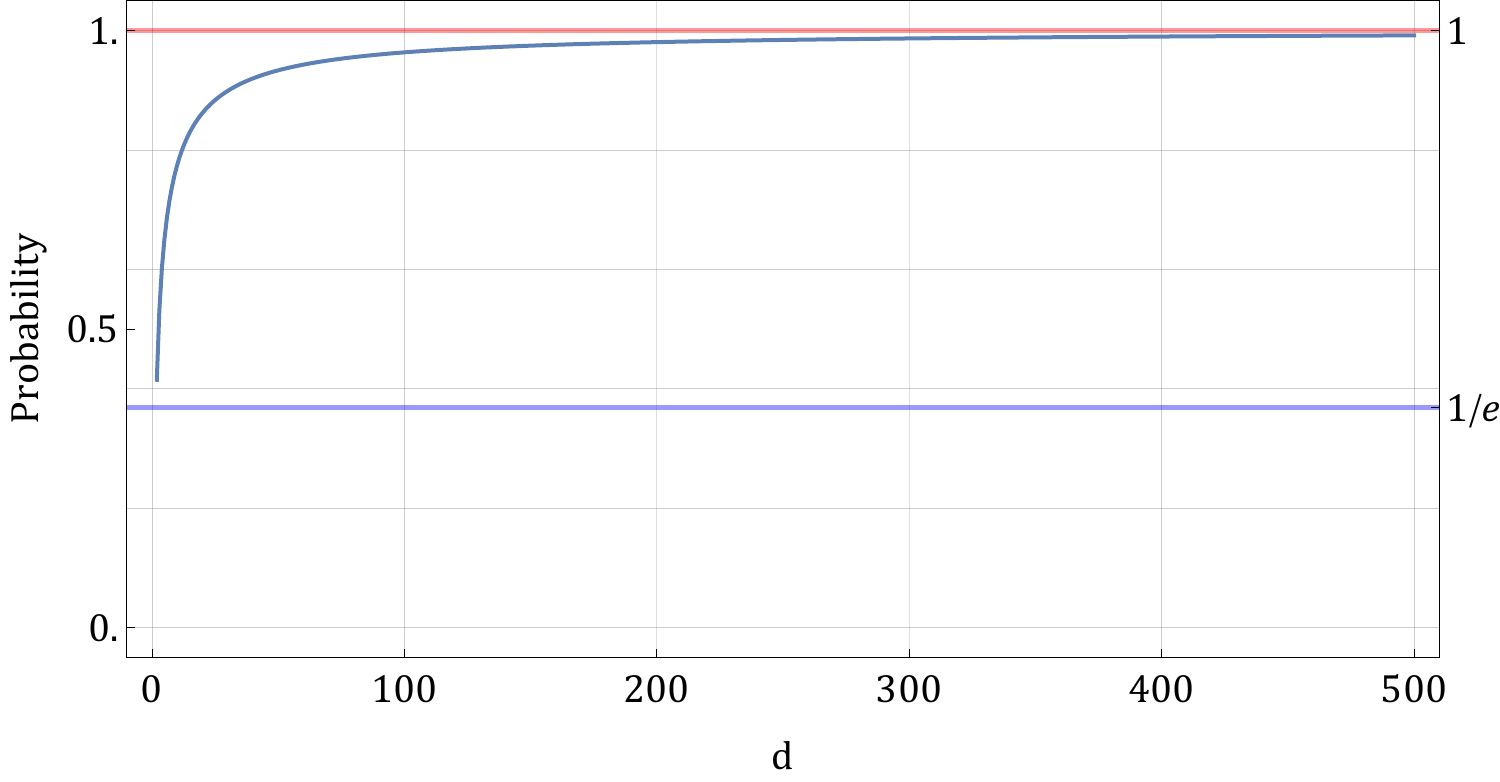}
		\caption{
			Results of the numerical optimization for the the probability of the $\aot^L$ sequence for the case $L=d+1$ or ${d=\DC(\aot^L)-1}$, for $d=2,\ldots,500$. The quantum value seems to converge to the trivial probability bound of $1$, and these models easily outperform the $1/e$ bound suggested by the classical scenario. See \appref{app:quantummodels} for the full result.}
		\label{fig:quantum-ot}
	\end{figure}

	For large $d$, the optimal probability is highly sensitive on the value of $\theta_0$, see \appref{app:quantummodels}. A good performance, however, is given by the angle $\theta_0 = (2\pi/d)(1-1/d)$, suggesting that it is asymptotically optimal, even if for small $d$ some deviation from the true optimum can be found.	To guess this particular value, we analyzed the action of the unitary $U_0$. 
	We recall that applying $d$ times the unitary $U_0$ gives
	\begin{equation}
	\begin{split}
	U_0^d = \e^{-\mathrm{i} \theta_0 d H } =  \e^{-\mathrm{i}\theta_0 d F \left(\sum_{k=0}^{d-1} k \ketbra{k}\right) F^\dagger }\\
	= F \left(\sum_{k=0}^{d-1} e^{-\mathrm{i}\theta_0 d k} \ketbra{k} \right) F^\dagger,  
	\text{ with  }[\,F\,]_{jk} = \frac{1}{\sqrt{d}} \e^{\frac{2 \pi i}{d} j k}.
	\end{split}
	\end{equation}
	This gives a transition probability between the last and the first state
	\begin{equation}
	\begin{split}
	= \left| \bra{d-1}U_0^d\ket{0} \right|^2 = \left| \sum_{k=0}^{d-1} [F]_{d-1,k} \, \e^{-i k d \theta_0} \, [F^\dagger]_{k,0} \right|^2 \\
	= \frac{1}{d^2} \left| \sum_{k=0}^{d-1} \exp\left[\frac{2 \pi i}{d} (d-1) k - i k d \theta_0 \right] \right|^2.
	\end{split}
	\end{equation}
	The maximum occurs when all the amplitudes are in phase, which implies the slowest rotating amplitude ($k = 1$) has to reach zero phase, i.e., when $\frac{2 \pi}{d} (d-1) - d \theta_0 = 0$. This gives $\theta_0 = (2\pi/d)(1-1/d)$, as desired. The POVM $E_0$ has been ignored here, which is at the origin of the deviation for small $d$. However, from the result of numerical calculation, see Fig.~\ref{fig:quantum-ot}, it seems that our approximation is good enough. See \appref{app:quantummodels} for details.
	
	This behavior is in close parallel with the optimal probability of transitioning forward in the classical case, $d/(d+1)$, which can be interpreted as the automaton transitioning $d$ states in $d+1$ steps. In the quantum case, the angle corresponds to a phase which synchronizes the amplitudes of all states when transitioning to the last state after $d$ steps. More details and intuitions on the quantum case are presented in \appref{app:quantummodels}.
	
	The results of this numerical optimization, shown in Fig.~\ref{fig:quantum-ot}, suggest that in the quantum case probability $1$ can be asymptotically reached. Of course, this fact cannot be proven rigorously by these numerical optimization methods, as any concrete computation, in a given dimension, results in a probability value strictly smaller than $1$.  To approach such questions a more sophisticated construction of quantum models is warranted and shall be the topic of future research.
	
	\section{Conclusions and outlook}\label{sec:conclusions}
	
	We explored optimal deterministic and probabilistic realization of a binary sequence with classical and quantum resources in the framework of finite-state automata. For deterministic realizations, where the difference between classical and quantum systems play no role, we introduced the notion of deterministic complexity quantifying the minimal dimension required, and provided an algorithm to efficiently compute it, as well as characterized some of its properties as a complexity measure. In the probabilistic case, i.e., $d < \DC$, we explore optimal realizations of sequences up to length $10$ via the Adam algorithm. This investigation improved a previously known model for the one-tick sequence probability with the discovery of the enhanced multicyclic models. Moreover, our results suggest that for any given sequence $\va$ and any dimension $d<\DC(\va)$ the optimal EMCM for the sequence $\aot^{\DC(\va)}$ provides an upper bound for the probability of the original sequence. In particular, by studying the optimal probability for all EMCMs, one can derive a universal upper bound of $1/e$ for all classical probabilistic realization ($d < \DC$) of a given sequence. 
	
	In the quantum case, we show how even simple models, with a small amount of memory resources, are already able to violate such universal classical bound, and consequently to outperform any classical scenario in which the realization of a sequence must necessarily be probabilistic. Moreover, the analysis of the performance of such quantum models for the one-tick sequences in high dimension, i.e., up to $d=500$, suggests that no analogous of such a nontrivial universal bound exists in the quantum case.
	
	Notwithstanding two fundamental limitations of our approach, namely, the analysis of only a finite set of sequences and the use of gradient descent methods, which for nonconcave problems guarantee only local maxima, we believe our findings provide a strong support for our conjectures. In fact, the number of sequences analyzed, counting also their realizations in different dimensions, was of the order of tens of thousands, all showing a similar structure of the optimal solution. It seems unlikely that any additional properties of sequences and models would emerge, which could invalidate these bounds. At the same time, an indication of the good performance of the gradient descent methods (in particular, Adam algorithm~\cite{Adam}) is that it was able to find very sparse solutions, containing mostly $0$s and $1$s.  
	For the quantum bound, similarly to the classical case, it would be very surprising if the limit of the one-tick sequence for $L=d+1$ and $d\rightarrow \infty$ would converge to some value strictly smaller than one.
	
	Our results stimulate future research in several directions. First, it would be interesting to understand whether the general upper bound based on the EMCM fails for longer sequences or if it is valid in general. This will also have consequence for the universal upper bound for lossy compression of a model.
	
	Without proof of its optimality, it may be possible that for certain longer sequences a nontrivial model may outperform its respective EMCM. The optimality of EMCMs themselves for one-tick sequences is also only conjectured at this time, but search for more general models have shown no evidence that EMCMs can be outperformed by uneven cycle lengths or independent probabilities, as described in \appref{app:otseqs}
	
	Moving away from the one-tick sequence, the optimal models for general sequences have proven to be exceptionally complex, showing a rich nontrivial behavior (e.g. \autoref{fig:pa_model} and \autoref{fig:neardc}). Despite this, nearly all optimal models found by our numerical search seem to fall into a small number of equivalence classes with same probability, despite the sequences having various lengths and dimensions, as shown in \autoref{fig:emcm_upperbound_l10} and \autoref{fig:emcm_upper_bounds}. The origin of these results is as of yet unknown, but suggest that tighter bounds on the conjecture may be established by exploiting additional structures of the sequences beyond their deterministic complexity.
	
	Our results for the quantum case also indicate that we may restrict our attention to instruments defined by single Kraus operators, significantly reducing the search space in an optimization. The question remains open of whether there are sequences and non-trivial dimensions for which classical and quantum lossy compression are equally efficient, i.e., if there exist Leggett-Garg-type temporal inequalities that are not violated by quantum systems, in analogy to the \emph{Guess-Your-Neighbor's-Input} non-local game~\cite{almeida2010}. Our results seem to suggest such scenarios are possible (e.g., see \autoref{fig:quantumperf2}), but this requires further investigation.
	
	%	In fact, in the quantum case our search is not able to find violation of the classical bound in several cases. In particular, the ansatz for the quantum model, based on the ideas presented in Ref.~\cite{Budroni2020}, seems to work well only for some sequences, and only for $d = \DC(\va) - 1$. The question remains open of whether there are sequences and dimensions for which classical and quantum lossy compression are equally efficient. In other words, if there exist Leggett-Garg-type temporal inequalities that are not violated by quantum systems.
	
	It is important to remark that the study of temporal correlations in the framework of finite-state automata is relatively recent, and few techniques have been so far developed to attack this problem. The numerical results and conjectures proposed in the present paper indicate new open problems and research directions that could stimulate the development on new intuitions and methods to investigate classical and quantum temporal correlations.
	
	Finally, the results and techniques utilized here could be generalized to arbitrary output alphabets, as well as the inclusion of inputs, which may reveal new information processing applications for finite automata models. All of these questions will be the subject of future research.
	
	\acknowledgements
	We thank \"Amin Baumeler, Cornelia Spee, and Giuseppe Vitagliano for useful discussions and comments on the manuscript. This work is supported  by the Austrian Science Fund (FWF) through projects ZK 3 (Zukunftskolleg) and F7113 (BeyondC).

	\appendix
	\begin{widetext}
		\appendixpage
		
		\section{Deterministic complexity and generating patterns}\label{app:dcpatterns}
		
		We start by proving that $\aot^L$ has $\DC=L$. The proof is based on a general criterion presented in \cite{SpeeNJP2020}. In simple terms, we may restate it here by saying that two steps of a sequence, generated by a minimal DFA, are associated with the same internal state if they ``have the same future'', i.e., if they generate the same sequence. For the $\aot^L$, the first state has the future ``emit a $1$ exactly after $L$ steps'', the second state ``emit $1$ exactly after $L-1$ steps'', and so on. Since there are no repetitions, it follows that each step of the sequence must be associated with a distinct state, giving $\DC=L$.
		
		As we have seen, the DC of a sequence $\va$ is computed by finding the minimal DFA which encodes its symbols utilizing a ``tail'' sequence leading to a ``cycle''. Such a construction can be succinctly specified with a \emph{pattern}, such as $\mathtt{00(101)}$, where the subsequence inside parentheses corresponds to a subsequence that must occur at least once in full, but may also repeat with possible truncations. The length of such patterns (in terms of $0$s and $1$s) corresponds to the desired DC. Thus, it suffices to find the minimum such patterns for a given sequence. However, patterns may not be unique, e.g. $\va = (0,0,1,0,1,0,1,1,0,1)$ can be written as $\mathtt{00(10101)}$ or $\mathtt{0010(101)}$, but all compatible and minimal patterns are of the same length.
		
		As pointed out, the ``tail'' part may have zero length if the entire sequence is a (possibly truncated) repeating pattern, as in $\va = (0,1,0,0,1,0,0,1,0,0,1)$ giving us the pattern $\mathtt{(010)}$, where the truncation may be observed in the last $0$ of the $4^\text{th}$ cycle. Under such analysis, one-tick sequences (and their $0 \leftrightarrow 1$ symmetric counterparts) are identified as the unique sequences saturating $\DC(\aot^L) = L$ for all sequences of length $L$. These sequences also have the maximum number ($L$) of potential patterns, as for example: $\aot^{5} = (0,0,0,0,1)$ gives $\mathtt{(00001)} \cong \mathtt{0(0001)} \cong \mathtt{00(001)} \cong \mathtt{000(01)} \cong \mathtt{0000(1)}$. \autoref{alg:dcpatterns}, which runs at $O(L^3)$, can be used to efficiently compute a sequence's deterministic complexity, as well as finding all of its minimal patterns. Notice that the algorithm is not restricted to a binary alphabet, and works the same way for an arbitrary one.
		
		\begin{algorithm}[H]
			\caption{Deterministic Complexity and patterns (DCPatterns). The idea is to assume the sequences have the form (tail)+(cycle), with respective lengths $l_1$ and $l_2$, such that $\DC(\va) = l_1 + l_2$. We thus test all such patterns counting up from DC, which ensures we find the minimal-state representation as early as possible.}\label{alg:dcpatterns}
			\begin{algorithmic}[1]
				\Procedure{DCPatterns}{$\va$}
				\State $L \gets \text{Length}(\va)$
				\State $patterns \gets \{\}$\Comment{We start with an empty list of patterns.}
				\State $dc \gets 0$ \Comment{We count from the bottom up in DC, stopping as early as possible.}
				\State $found \gets \mathrm{False}$ \Comment{When we find any patterns, we can stop.}
				\While{\textbf{not} $found$}
				\State $dc \gets dc + 1$
				\For{$l_1$ \textbf{in} $0, \dots, dc$} \Comment{For each possible tail length}
				\State $l_2 \gets dc - l_1$ \Comment{... assume the rest is a cycle.}
				\State $match \gets \mathrm{True}$ \Comment{Assume this is a valid pattern.}
				\For{$i$ \textbf{in} $0, \dots, (L - dc)$} \Comment{For every other symbol beyond DC}
				\If{$\va[dc+i] \neq \va[l_1 + (i \; \mathrm{mod} \; l_2)]$} \Comment{we test whether it can be extrapolated.}
				\State $match \gets \mathrm{False}$ \Comment{If it can't, this pattern fails and we stop here.}
				\State \textbf{break}
				\EndIf
				\EndFor
				\If{$match$} \Comment{If the pattern matches the entire sequence...}
				\State $patterns \gets patterns + \{(l_1,l_2)\}$\Comment{... we add the tuple $(l_1,l_2)$ to the list of patterns.}
				\State $found \gets \mathrm{True}$ \Comment{We stop at this $dc$, but continue searching for patterns.}
				\EndIf
				\EndFor
				\EndWhile
				\State \textbf{return} $dc, patterns$\Comment{Return the optimum $dc$ and the list of patterns}
				\EndProcedure
			\end{algorithmic}
		\end{algorithm}
		
		\subsection{The structure of patterns}\label{app:patterns}
		
		Arbitrary patterns are not necessarily unique or minimal in the number of states. As a concrete example, the patterns $\mathtt{00(00)}$ and $\mathtt{01(0101)}$, of lengths 4 and 6 respectively, are equivalent to the minimal patterns $\mathtt{(0)}$ and $\mathtt{(01)}$, of lengths 1 and 2.
		
		Interestingly, the space of unique minimal (binary) patterns using exactly $d$ states can be directly related to the unique minimal unary deterministic finite automata (uDFAs) with exactly $d$ states. An analysis and enumeration of these uDFAs has been presented by Nicaud~\cite{Nicaud1999} and Domaratzki \etal~\cite{Domaratzki2002}. For $d = 1, 2, \dots$ there are $N_\text{uDFA}(d) = 2,\ 4,\ 12,\ 30,\ 78,\ 180,\ 432, \dots$ such uDFAs, and thus unique minimal patterns with exactly $d$ states.
		
		The argument for uDFAs can be easily adapted to our case and can be understood as follows. For a pattern of length $\ell$ to be minimal, it must satisfy two obvious conditions: $(i)$ the tail is minimal, and $(ii)$ the cycle is minimal.
		
		Suppose the ``tail'' and ``cycle'' parts of the pattern are given by the strings (or \emph{words}) $t$ and $c$, respectively, and let $tc$ be their concatenation.
		For condition $(i)$ to be valid, the last symbol of $c$ must be different from the last symbol of $t$, otherwise, one could include the last transition of the tail into the cycle while removing the last transition of the cycle, obtaining a pattern of length $\ell - 1$ generating the same infinite sequence. For instance, the pattern $\mathtt{01(001)}$ is not minimal because it could be reduced to $\mathtt{0(100)}$.
		
		Condition $(ii)$ implies $c$ is a so-called \emph{primitive word}~\cite{allouche2003automatic}, i.e., it must be non-empty and not be expressible as $c = w^m$ for $w$ a smaller word and $m \in \mathbb{N}$. As a concrete example, $\mathtt{0101}$ is not a primitive word of length $4$, as it is a repetition of a shorter word $\mathtt{01}$ of length $2$. Clearly, the number of primitive words of length $n$ relates to the divisors $d$ of $n$. For a given alphabet of $k$ symbols, the number of primitive words $\psi_k(n)$ of length $n$ may be computed~\cite{Nicaud1999,Domaratzki2002,allouche2003automatic} in terms of the M\"obius function $\mu(d)$:
		\begin{equation}
		\mu(d) = \begin{cases}
		0, & \text{if}\ d\ \text{is divisible by some}\ x^2 > 1,\ \text{with}\ x \in \mathbb{N}; \\
		(-1)^s, & \text{if}\ d = p_1 p_2 \dots p_s, \ \text{where}\ p_i \ \text{are distinct primes}.
		\end{cases}
		\end{equation}
		The number of primitive words is then given by
		\begin{equation}\label{eq:psik}
		\psi_k(n) = \sum_{d | n} \mu(d) k^{n/d}.
		\end{equation}
		The expression for the number of minimal patterns of length $\ell$ over a $k$-symbol alphabet is, thus,
		\begin{equation}\label{npatt}
		N_k(\ell) = \psi_k(\ell) + \sum_{i=1}^{\ell-1} (k-1) k^{i-1} \psi_k(\ell-i),
		\end{equation}
		where the term $\psi_k(\ell)$ counts the number of length-$\ell$ cycles (i.e., case of no tail), whereas $(k-1) k^{i-1} \psi_k(\ell-i)$ counts the number of length-$(\ell-i)$ cycles together with length-$i$ tails, where one element of the tail is constrained by condition $(i)$ above, giving the $(k-1) k^{i-1}$ factor.
		
		\section{Optimal classical models for one-tick sequences}\label{app:otseqs}
		
		Given their special role in the problem at hand, the optimal values for one-tick sequences deserve a more detailed analysis. In Ref.~\cite{Budroni2020}, a class of models was proposed for the optimal bounds for the one-tick sequence. We have since found a further generalization which gives higher values in certain scenarios. Due to their sparse structure, these models are unusually difficult to obtain through typical unconstrained numerical optimizations. Given their significantly higher performance we conjecture these models to be optimal.
		
		\subsection{One-way, cyclic and multicyclic models}\label{app:oldclockmodels}
		
		The model presented in Ref.~\cite{Budroni2020} was called the {\it  multicyclic model} which included as special cases the {\it one-way model} and the {\it cyclic model}.
		
		The one-way model was found to be generally best for $d = \DC(\aot^L) - 1 = L-1$, whereas the (multi)cyclic models were found to outperform it for certain $d < \DC(\aot^L) - 1$. The multicyclic model divides the states into equal sub-cycles, within each cycle transitions occur deterministically, except in the last state of each sub-cycle, where the cycling back or the transitioning to the next cycle (or even the emission of output $1$ for the last sub-cycle) is probabilistic. These restrictions imply that the total number of cycles $n$ and the size of the cycles $k$ must obey $d = n k$.
		
		\begin{figure}[H]\centering
			\includegraphics[width=0.52\linewidth]{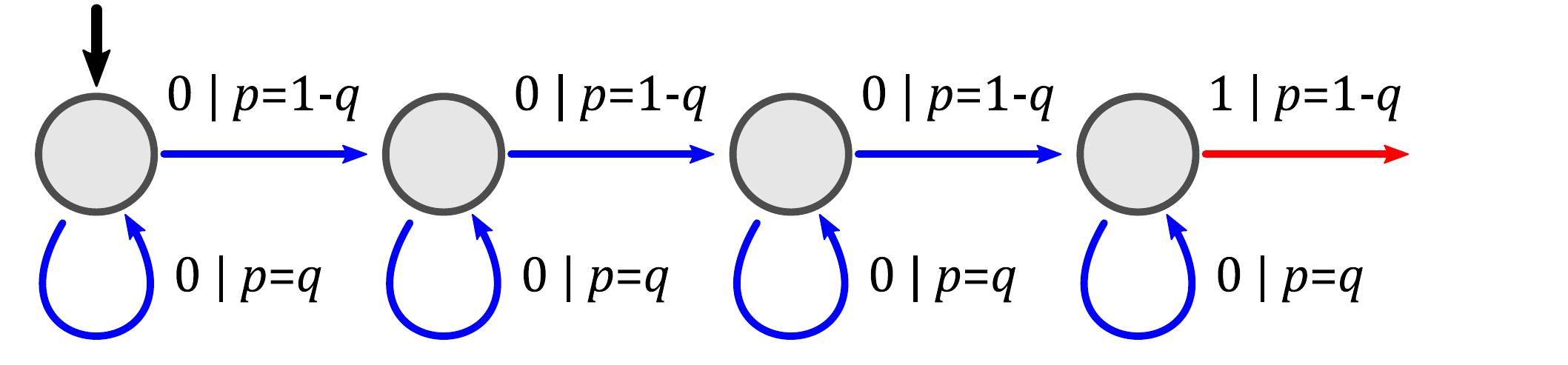}\\
			
			(a)
			
			\includegraphics[width=0.52\linewidth]{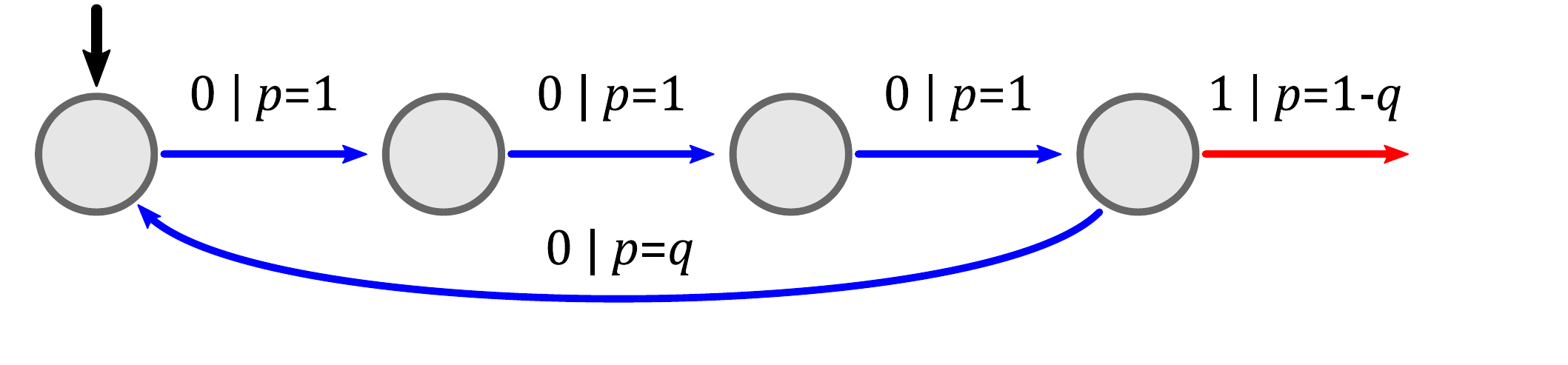}\\
			
			(b)
			
			\includegraphics[width=0.52\linewidth]{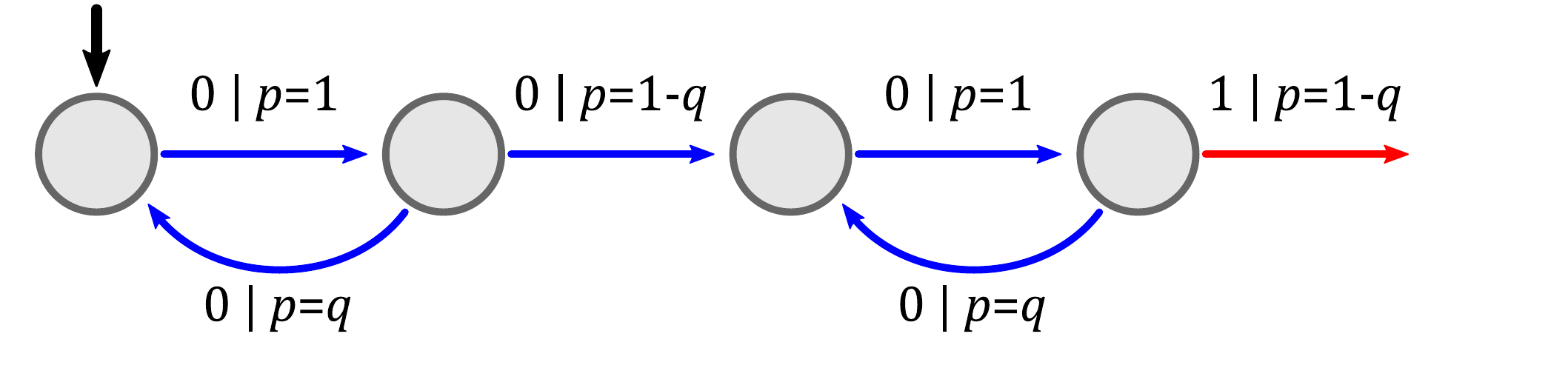}\\
			
			(c)
			\caption{Different models for the one-tick sequences in $d = 4$: (a) one-way model, (b) cyclic model, (c) multicyclic models. In the one-tick sequence, the state after the final transition on outcome 1 is irrelevant.}
			\label{fig:clockmodels}
		\end{figure}
		
		\subsection{Generalized \& enhanced multicyclic models}\label{app:EMCM}
		
		\begin{figure}[H]\centering
			\includegraphics[width=0.6\linewidth]{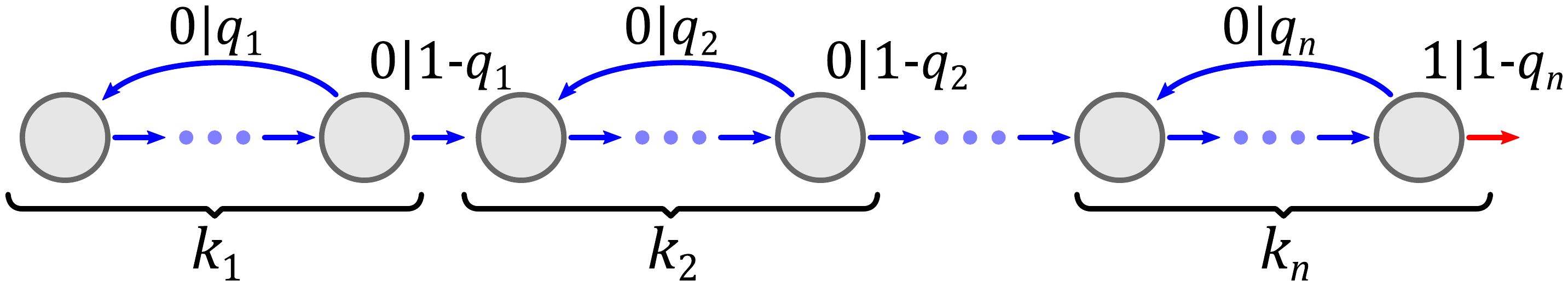}
			\caption{The structure of generalized multicyclic models, which consist of splitting the $d$ states into $n$ blocks of various sizes $k_i$, each block corresponding to an internally-deterministic cycle with independent probability $q_i$ of cycling, and $1-q_i$ of moving to the next cycle. Only the last state of the last block has a probability $1-q_n$ of outputting $1$. All optimal models for the one-tick sequence investigated are of this form.}
			\label{fig:gmcm}
		\end{figure}
		
		In this section, we explain the origins of the \emph{enhanced multicyclic model} (EMCM) as a generalization of the models introduced in Ref.~\cite{Budroni2020}.
		
		The multicyclic model assumed $d = nk$, but if $d$ and $k$ are coprime, then either $k = 1$ or $k = d$, and we obtain the one-way or cyclic models, respectively, which can be sub-optimal. To fully generalize such models we must allow any possible decomposition of the $d$ states into blocks, whose cycle probabilities may be optimized independently.
		
		In such a \emph{generalized multicyclic model} (GMCM), the $T_0$ transition matrix consists of $n$ cycles with internal deterministic transitions, and a probabilistic transition to the initial state of the cycle, with probability $q_i$, or to the first state in the next cycle, with probability $1-q_i$, for the $i$-th cycle. Once the last state is reached, we have a nonzero probability, i.e., $1-q_n$, for the output $1$. An explicit example of $T_0$ is shown in Eq.~\eqref{eq:TGMCM} for a $d=9$ model, and a general structure in shown in \autoref{fig:gmcm}. The $T_1$ matrix is $T_1={\rm diag}(0,\ldots,0,1-q_n)$. Notice that since the last transition is irrelevant, in $T_1$ we could have chosen any other position in the last row for the non-zero entry $1-q_n$.
		
		\begin{equation}\label{eq:TGMCM}
		T_0 = \left[
		\begin{array}{cccc|ccc|cc}
		0 & 1 & 0 & 0 & 0 & 0 & 0 & 0 & 0 \\
		0 & 0 & 1 & 0 & 0 & 0 & 0 & 0 & 0 \\
		0 & 0 & 0 & 1 & 0 & 0 & 0 & 0 & 0 \\
		q_1 & 0 & 0 & 0 & 1-q_1 & 0 & 0 & 0 & 0 \\ \hline
		0 & 0 & 0 & 0 & 0 & 1 & 0 & 0 & 0 \\
		0 & 0 & 0 & 0 & 0 & 0 & 1 & 0 & 0 \\
		0 & 0 & 0 & 0 & q_2 & 0 & 0 & 1-q_2 & 0 \\ \hline
		0 & 0 & 0 & 0 & 0 & 0 & 0 & 0 & 1 \\
		0 & 0 & 0 & 0 & 0 & 0 & 0 & q_3 & 0 \\
		\end{array}
		\right]
		\end{equation}
		
		Such a GMCM is defined by several parameters: a \emph{signature} $\vec{k}=(k_1,\ldots,k_n)$ denoting the size of each of the $n$ cycles, with $k_i\in \mathbb{N}^+$ and $\sum_i k_i=d$, and $\vec{q}=(q_1,\ldots, q_n)$ denoting the probability of each cycling transition to occur, with $0\leq q_i< 1$ (the strict inequality is important as having one of the $q_i=1$ would give probability zero for the one-tick sequence). 
		
		Note that there are in total $2^{d-1}$ potential signatures for a given $d$, corresponding to the number of integer compositions\footnote{Also known as ordered partitions. For example, $d=4$ has 8 compositions: $(1,1,1,1)$, $(1,1,2)$, $(1,2,1)$, $(2,1,1)$, $(2,2)$, $(1,3)$, $(3,1)$, and $(4)$. These compositions can be easily enumerated.} of the number $d$. A priori, none of these could have been ruled out, so an exhaustive numerical survey of these models was performed for $L = 3, \dots, 10$ and $d = 1, \dots, L-1$. The optimal models found are shown in \autoref{tbl:gmcmdata}, where a few structures became apparent: due to optimal models having the same probabilities for the same block sizes, signatures largely obey permutation symmetry, but favoring bigger blocks at the beginning. 
		
		\begin{table}[H]
			\renewcommand{\arraystretch}{1.5}
			\setlength{\tabcolsep}{0.75em}
			\centering
			\begin{tabular}{|c|c|c|} \hline
				\multicolumn{3}{|c|}{$L=3$} \\ \hline
				$d$ & $p(\aot^{L}|G,d)$ & $\vec{k}$ \\ \hline
				$1$ & $0.148148$ & $(1)_{0}$ \\
				$2$ & $0.296296$ & $(1,1)_{0}$ \\ \hline 
			\end{tabular} \qquad \begin{tabular}{|c|c|c|} \hline
				\multicolumn{3}{|c|}{$L=4$} \\ \hline
				$d$ & $p(\aot^{L}|G,d)$ & $\vec{k}$ \\ \hline
				$1$ & $0.105469$ & $(1)_{0}$ \\
				$2$ & $0.250000$ & $(2)_{0}$ \\
				$3$ & $0.316406$ & $(1,1,1)_{0}$ \\ \hline 
			\end{tabular} \qquad \begin{tabular}{|c|c|c|} \hline
				\multicolumn{3}{|c|}{$L=5$} \\ \hline
				$d$ & $p(\aot^{L}|G,d)$ & $\vec{k}$ \\ \hline
				$1$ & $0 .081920$ & $(1)_{0}$ \\
				$2$ & $0 .148148$ & $(2)_{1}$ \\
				$3$ & $0 .250000$ & $(1,2)^{*}_{0}, (3)_{1}$ \\
				$4$ & $0 .327680$ & $(1,1,1,1)_{0}$ \\ \hline 
			\end{tabular}
			\\[1em]
			\begin{tabular}{|c|c|c|} \hline
				\multicolumn{3}{|c|}{$L=6$} \\ \hline
				$d$ & $p(\aot^{L}|G,d)$ & $\vec{k}$ \\ \hline
				$1$ & $0.066980$ & $(1)_{0}$ \\
				$2$ & $0.148148$ & $(2)_{0}$ \\
				$3$ & $0.250000$ & $(3)_{0}$ \\
				$4$ & $0.296296$ & $(2,2)_{0}$ \\
				$5$ & $0.334898$ & $(1,1,1,1,1)_{0}$ \\ \hline 
			\end{tabular} \; \begin{tabular}{|c|c|c|} \hline
				\multicolumn{3}{|c|}{L=7} \\ \hline
				$d$ & $p(\aot^{L}|G,d)$ & $\vec{k}$ \\ \hline
				$1$ & $0.056653$ & $(1)_{0}$ \\
				$2$ & $0.105469$ & $(2)_{1}$ \\
				$3$ & $0.148148$ & $(1,2)^{*}_{0},\; (3)_{2}$ \\
				$4$ & $0.250000$ & $(1,3)^{*}_{0},\; (4)_{1}$ \\
				$5$ & $0.296296$ & $(1,2,2)^{*}_{0}$ \\
				$6$ & $0.339917$ & $(1,1,1,1,1,1)_{0}$ \\ \hline 
			\end{tabular} \; \\[1em] \begin{tabular}{|c|c|c|} \hline
				\multicolumn{3}{|c|}{L=8} \\ \hline
				$d$ & $p(\aot^{L}|G,d)$ & $\vec{k}$ \\ \hline
				$1$ & $0.049087$ & $(1)_{0}$ \\
				$2$ & $0.105469$ & $(2)_{0}$ \\
				$3$ & $0.148148$ & $(3)_{1}$ \\
				$4$ & $0.250000$ & $(4)_{0}$ \\
				$5$ & $0.250000$ & $(1,1,3)^{*}_{0},\; (2,3)^{*}_{0},\; (4,1)_{1},\; (5)_{2}$ \\
				$6$ & $0.316406$ & $(2,2,2)_{0}$ \\
				$7$ & $0.343609$ & $(1,1,1,1,1,1,1)_{0}$ \\ \hline 
			\end{tabular}
			\\[1em]
			\begin{tabular}{|c|c|c|} \hline
				\multicolumn{3}{|c|}{L=9} \\ \hline
				$d$ & $p(\aot^{L}|G,d)$ & $\vec{k}$ \\ \hline
				$1$ & $0.043305$ & $(1)_{0}$ \\
				$2$ & $0.081920$ & $(2)_{1}$ \\
				$3$ & $0.148148$ & $(3)_{0}$ \\
				$4$ & $0.148148$ & $(3,1)_{1},\; (4)_{3}$ \\
				$5$ & $0.250000$ & $(1,4)^{*}_{0},\; (5)_{1}$ \\
				$6$ & $0.296296$ & $(3,3)_{0}$ \\
				$7$ & $0.316406$ & $(1,2,2,2)^{*}_{0}$ \\
				$8$ & $0.346439$ & $(1,1,1,1,1,1,1,1)_{0}$ \\ \hline 
			\end{tabular} \qquad \begin{tabular}{|c|c|c|} \hline
				\multicolumn{3}{|c|}{L=10} \\ \hline
				$d$ & $p(\aot^{L}|G,d)$ & $\vec{k}$ \\ \hline
				$1$ & $0.038742$ & $(1)_{0}$ \\
				$2$ & $0.081920$ & $(2)_{0}$ \\
				$3$ & $0.105469$ & $(3)_{2}$ \\
				$4$ & $0.148148$ & $(4)_{2},\; (1,3)^{*}_{0}$ \\
				$5$ & $0.250000$ & $(5)_{0}$ \\
				$6$ & $0.250000$ & $(1,1,4)^{*}_{0},\; (2,4)^{*}_{0},\; (5,1)_{1},\;  (6)_{2}$ \\
				$7$ & $0.296296$ & $(1,3,3)^{*}_{0}$ \\
				$8$ & $0.327680$ & $(2,2,2,2)_{0}$ \\
				$9$ & $0.348678$ & $(1,1,1,1,1,1,1,1,1)_{0}$ \\ \hline 
			\end{tabular}
			\caption{All optimal generalized multicyclic models ($G$) for $3 \le L \le 10$, with their probabilities and signatures $\vec{k}$. Subscripts on signatures indicate the optimal initial state $z$ (starting from $0$), and an asterisk indicates that all permutations of the signature were found to be equivalent. Several non-uniform optimal models appear, improving upon results in Ref.~\cite{Budroni2020}.}
			\label{tbl:gmcmdata}
		\end{table}
		
		Despite the fact that many signatures of the optimal models found specify non-uniform cycle lengths (e.g. $(1,2,2)$), in all such cases the smallest cycles found had $q_i = 0$, that is, they do not behave like cycles at all, but instead acting as a series of deterministic transitions. This renders the models given by $(1,1,4)$ and $(2,4)$ for $L=10$ identical, for instance. Furthermore, only the largest cycles required $q_i > 0$, and in all cases the optimal models display $q_i = q_j$ if $k_i = k_j$. Finally, in all cases where a signature did not satisfy permutation symmetry, the optimal model contained the deterministic transitions at the end.
		
		This suggests that we can always permute the model so that all deterministic transitions occur at the end, as a single deterministic block, while all identical probabilistic cycles occur at the beginning, thus unifying all such optimal models within the same structure. 
		
		This inspires us to the following simplified model: we only need to consider models of the form $\vec{k} = (k, \dots, k, t)$, where $d = n k + t$, with a number $n$ of $k$-sized blocks, i.e., the vast majority of signatures can be dismissed. In such models, we always have $n = \floor{d/k}$, and $t = d - n k$, and an initial state $z = 0, \dots, k-1$.
		
		These are the enhanced multicyclic models (EMCMs), as discussed in the main text, which we conjecture are optimal. They can be fully specified by the $5$-tuple of parameters $(L,n,k,t,z)$. \autoref{tbl:emcmparams} details the parameters describing various optimal models.
		
		In addition to the comprehensive low-dimension survey of GMCMs which led us to the EMCMs, we have also investigated some higher-dimensional cases up to $d = 20$, and $20 < L \le 50$ using both GMCMs and the general unrestricted models, as discussed in \appref{app:survey}. In all cases, the best model found was always in the form of an EMCM, which further supports our claim that EMCMs are optimal.
		
		\begin{table}[H]
			\footnotesize
			\renewcommand{\arraystretch}{1.5}
			\setlength{\tabcolsep}{0.75em}
			\centering
			\begin{tabular}{|c|c|c|c|c|c|c|c|c|c|} \hline
				\diagbox{L}{d} & $1$ & $2$ & $3$ & $4$ & $5$ & $6$ & $7$ & $8$ & $9$ \\ \hline
				$2$ & $(1,1,0,0)$ & $\;$ & $\;$ & $\;$ & $\;$ & $\;$ & $\;$ & $\;$ & $\;$ \\ \hline
				$3$ & $(1,1,0,0)$ & $(2,1,0,0)$ & $\;$ & $\;$ & $\;$ & $\;$ & $\;$ & $\;$ & $\;$ \\ \hline
				$4$ & $(1,1,0,0)$ & $(1,2,0,0)$ & $(3,1,0,0)$ & $\;$ & $\;$ & $\;$ & $\;$ & $\;$ & $\;$ \\ \hline
				$5$ & $(1,1,0,0)$ & $(1,2,0,1)$ & $(1,3,0,1)$ & $(4,1,0,0)$ & $\;$ & $\;$ & $\;$ & $\;$ & $\;$ \\ \hline
				$6$ & $(1,1,0,0)$ & $(1,2,0,0)$ & $(1,3,0,0)$ & $(2,2,0,0)$ & $(5,1,0,0)$ & $\;$ & $\;$ & $\;$ & $\;$ \\ \hline
				$7$ & $(1,1,0,0)$ & $(1,2,0,1)$ & $(1,3,0,2)$ & $(1,4,0,1)$ & $(2,2,1,0)$ & $(6,1,0,0)$ & $\;$ & $\;$ & $\;$ \\ \hline
				$8$ & $(1,1,0,0)$ & $(1,2,0,0)$ & $(1,3,0,1)$ & $(1,4,0,0)$ & $(1,5,0,2)$ & $(3,2,0,0)$ & $(7,1,0,0)$ & $\;$ & $\;$ \\ \hline
				$9$ & $(1,1,0,0)$ & $(1,2,0,1)$ & $(1,3,0,0)$ & $(1,4,0,3)$ & $(1,5,0,1)$ & $(2,3,0,0)$ & $(3,2,1,0)$ & $(8,1,0,0)$ & $\;$ \\ \hline
				$10$ & $(1,1,0,0)$ & $(1,2,0,0)$ & $(1,3,0,2)$ & $(1,4,0,2)$ & $(1,5,0,0)$ & $(1,6,0,2)$ & $(2,3,1,0)$ & $(4,2,0,0)$ &	$(9,1,0,0)$ \\ \hline
			\end{tabular}
			\caption{Optimal $(n,k,t,z)$ parameters for enhanced multicyclic models for various $L$ and $d$. In the case multiple sets of parameters resulted in the same probability, the set with the smallest $t$ was chosen to highlight the cases in which the EMCM structure is strictly required. Including $L$, the five parameters fully specify the probability for a given $(L,d)$.}
			\label{tbl:emcmparams}
		\end{table}
		
		\subsection{Reducible sequences and models}
		
		Enhanced multicyclic models can be intuitively understood as follows. The initial shift $z$, can be interpreted as ``increasing'' the sequence length (without affecting $d$): since the transitions within each cycle block are deterministic, starting from the state $z+1$ is equivalent to lengthening the sequence by $z$ steps and starting on the first state. Similarly, the final sequence of $t$ deterministic transitions do not alter the probability, defined by the cycle blocks. Thus, one may ``trim'' these deterministic transitions by removing $t$ steps in the sequence together with $t$ states. The result of both transformations is a multicyclic model starting at the first state, with the same probability as before.
		
		Finally, each block of size $k$ is equivalent to a single state of the one-way model of reduced dimension, effectively a time-scaling by a factor of $1/k$. In other words, the initial state $z$ and the final deterministic block $t$ both work together to perfectly synchronize the cycles with the number of transitions.
		
		In summary, the optimal probability for an EMCM with $d$ states and one-tick sequence of length $L$ can then easily be computed by performing the following transformations, which leave the probability unchanged:
		\begin{enumerate}
			\item For a given $L, d$, pick $k, z \in \mathbb{N}$ obeying $1 \le k \le d$ and $0 \le z \le k-1$. Compute $n = \floor{d/k}$ and $t = d - nk$.
			\item If $t > 0$, we may discard the entire final deterministic block, effectively reducing the dimension, if we also shorten the one-tick sequence by the same amount $t$ (removing $t$ $0$s). In the process, we incorporate the deterministic output $1$ into the final probabilistic transition forwards in the last state of the last cycle. In the above table, this corresponds to moving diagonally towards the top-left corner by $t$ cells.
			\item If the initial state is not the first state, i.e. $z > 0$, we may increase the sequence length by adding $z$ $0$s, as the transitions within cycle blocks are deterministic. This effectively shifts the initial state to the first state of the first block. This corresponds to moving down the table by $z$ cells.
			\item Once these transformations have been performed, we have transformed the problem into the $L \mapsto L-t+z$, $d \mapsto d-t$ scenario with a simple multicyclic model.
			\item Finally, since all blocks have the same length $k$ with $k-1$ deterministic transitions, we may divide both the new $L$ and $d$ by $k$, obtaining a one-way model for $L \mapsto L' = \frac{L - t + z}{k}$ and $d \mapsto d' = \frac{d - t}{k}$.
		\end{enumerate}
		
		We then have the model and sequence in an irreducible canonical form with an alternative $(L',d')$ of equal probability, and may directly apply the optimal probability for the one-way model, given by the negative binomial distribution
		\begin{equation}
		F_{\rm ow}(L,d)=\binom{L-1}{d-1} \left(1-\frac{d}{L}\right)^{L-d}\left(\frac{d}{L}\right)^d,
		\end{equation}
		where $q = 1 - d/L$ is the probability of self-transitions and $1-q=d/L$ the probability of forward transitions.
		
		The only free parameters in such an approach are finding the optimal $k$ and $z$, which require a straightforward numerical optimization on the order of $O(d^2)$.
		
		This optimality of enhanced multicyclic cycles for one-tick sequences, due to their reducibility, leads to an alternative physical interpretation of these models. The resulting uniform multicyclic models with cycle lengths $k$ represent a ``\emph{time-scaling symmetry}''. Since every $k-1$ transitions occur deterministically they may be considered as mere \emph{delays} which introduce no relevant dynamics to the behavior of the model. The result of this reduction is a ``time scaling'' by a factor of $1/k$, which produces an irreducible version of the model that captures all of its dynamics in the smallest number of transitions and states.
		
		\section{Numerical survey of optimal classical models}\label{app:survey}

		For arbitrary sequences $\va$ optimal models are far from obvious, and must be optimized numerically. In practice we tackle problem in Eq.~\eqref{eq:optprob} numerically, which can only provide us with a lower estimate $\Oest$ for the true classical upper-bound of a sequence, $\Ostar$.
		
		Instead of approaching this constrained problem as is, we convert it into the equivalent unconstrained problem by defining the real-valued $d \times $d matrices $B_0$ and $B_1$. We can convert these into $T_0$ and $T_1$ in the constrained problem via the normalization procedure
		\begin{equation}
		[T_a]_{ij} = \frac{[B_a]_{ij}^2}{\sum_\ell \left( [B_0]_{i\ell}^2 + [B_1]_{i\ell}^2 \right)}, \quad \text{for}\, a = 0, 1.
		\end{equation}
		
		We performed the optimization using the Adam algorithm~\cite{Adam} as implemented in the PyTorch package~\cite{pytorch}. For each sequence $\va$, 25 initial $B_a$ were chosen at random with entries drawn from the uniform distribution on the symmetric unit interval $[-1,1]$. We aimed for an accuracy of $10^{-8}$, which seemed reasonable given the typical numerical fluctuations observed, and an initial learning rate of $0.005$. The best probability from all samples obtained was taken as the optimum estimate for the bound for each sequence, and the optimal model $T$ found was stored for later analysis.
		
		We estimated the optimal models and probabilities $\Oest$ for all sequences of length $L = 3,\dots,10$ and $d = 1,2,\dots,\DC(\va)-1$. Since the models are invariant under exchange in output symbols ($0 \leftrightarrow 1$), we may focus on only half of the sequences by fixing $a_1 = 0$.
		
		The results of our optimizations are shown, in aggregate, in \autoref{fig:emcm_upper_bounds}, where we compare the probabilities found for all sequences with the optimal EMCM probabilities. The models obtained in our numerical survey show a rich and complex structure, with many nontrivial mixtures of deterministic and probabilistic transitions (e.g. \autoref{fig:pa_model}). \autoref{fig:neardc} displays the behavior of various $d = \DC(\va) - 1$ models.
		
		While a deeper understanding of such structures is subject of ongoing research, the fact all sequences perform worse than the one-tick sequence may be interpreted as follows: these results show that there is a nontrivial trade off occurring when models attempt to reach a higher probability with less resources. The way other sequences switch between $0$ and $1$ multiple times forces the memory resources to be spread over multiple incompatible transitions, leading to a worse performance overall.
		
		\begin{figure}[H]\centering
			\includegraphics[width=0.85\linewidth]{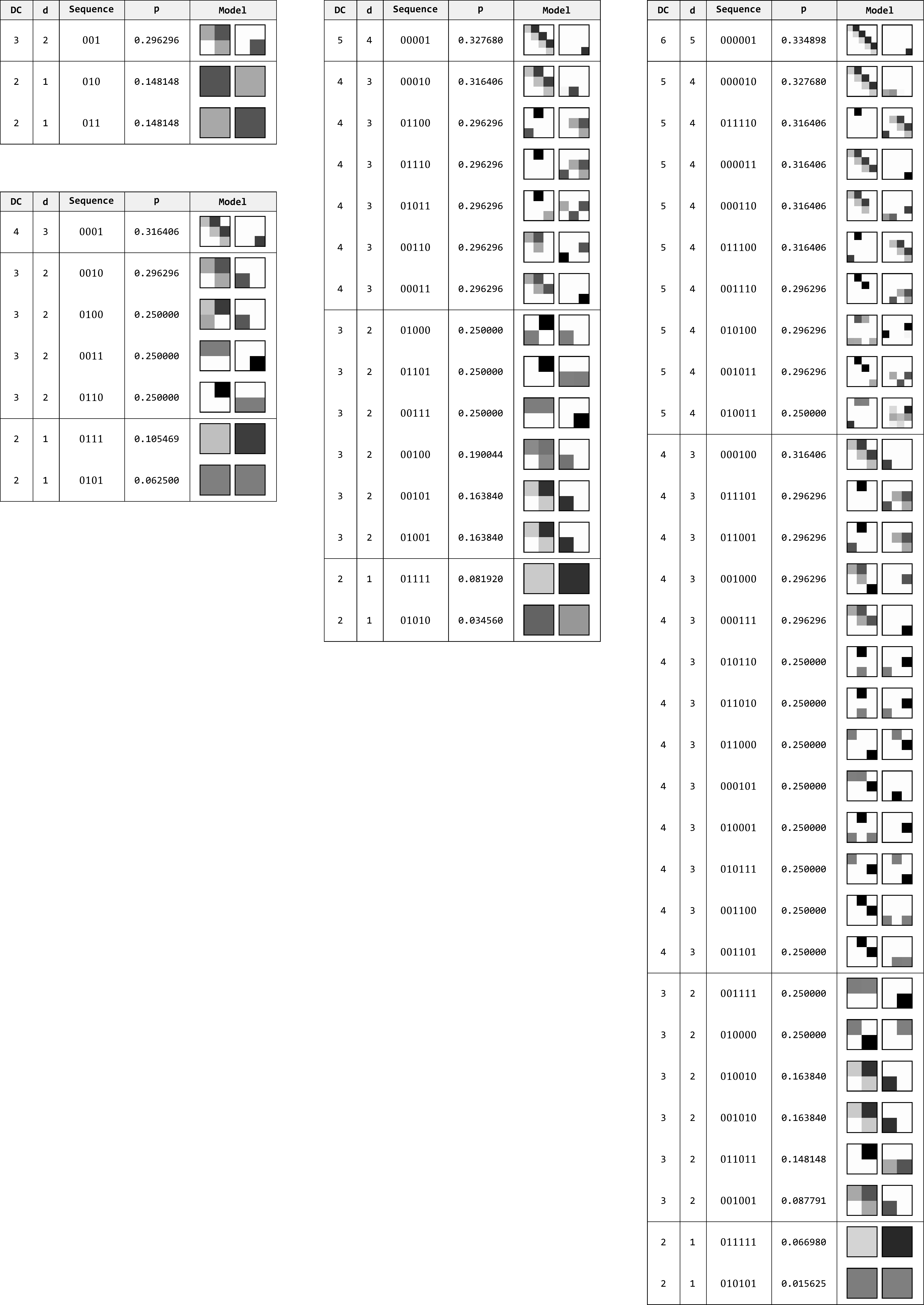}
			\caption{Optimal models found for $L = 3, 4, 5, 6$ sequences, grouped by decreasing DC, and sorted by decreasing probability. Only $d = \DC(\va) - 1$ cases are shown. The pictorial representation of the model matrices $(T_0,T_1)$ has $0$ entries as white, and $1$ as black, with intermediate values in gray. Despite the large variety in structure and sparsity of the models, nearly all sequences fall into similar equivalence classes with the same probabilities. The one-tick sequence (possibly truncated to DC) always outperformed the other sequences, acting as an upper-bound for any $(L,d)$ combination.}
			\label{fig:neardc}
		\end{figure}
		
		\section{Conjectured upper bounds}\label{app:conjecture}
		
		All the available data so far seems to confirm the conjecture that one-tick sequences may be used as upper estimates for optimal use of the $d$ states available. A graphical representation of our conjecture is shown in \autoref{fig:emcm_upper_bounds}, where it is clear that tighter estimates for the bound may still be available, by identifying additional structures in the sequences and the origin of the probability equivalence classes, which appear as ``plateaus''.
		
		If the conjecture holds and the EMCMs are indeed optimal, this would imply the existence of a universal maximum probability for any classical model with $d < \DC(\va)$, given by $1/\e$ (blue line at the top of the plots). This is the limiting probability of the $d = L-1$ one-way model for $L \to \infty$, given by $p = \lim_{L \to \infty} \left(1 - \frac{1}{L} \right)^L$.
		
		Given the promising results shown, it remains to be proven that the one-tick sequence gives the optimal usage of the $d$ states. This is the focus of ongoing research.
		
		\begin{figure}[H]\centering
			\includegraphics[width=1\linewidth]{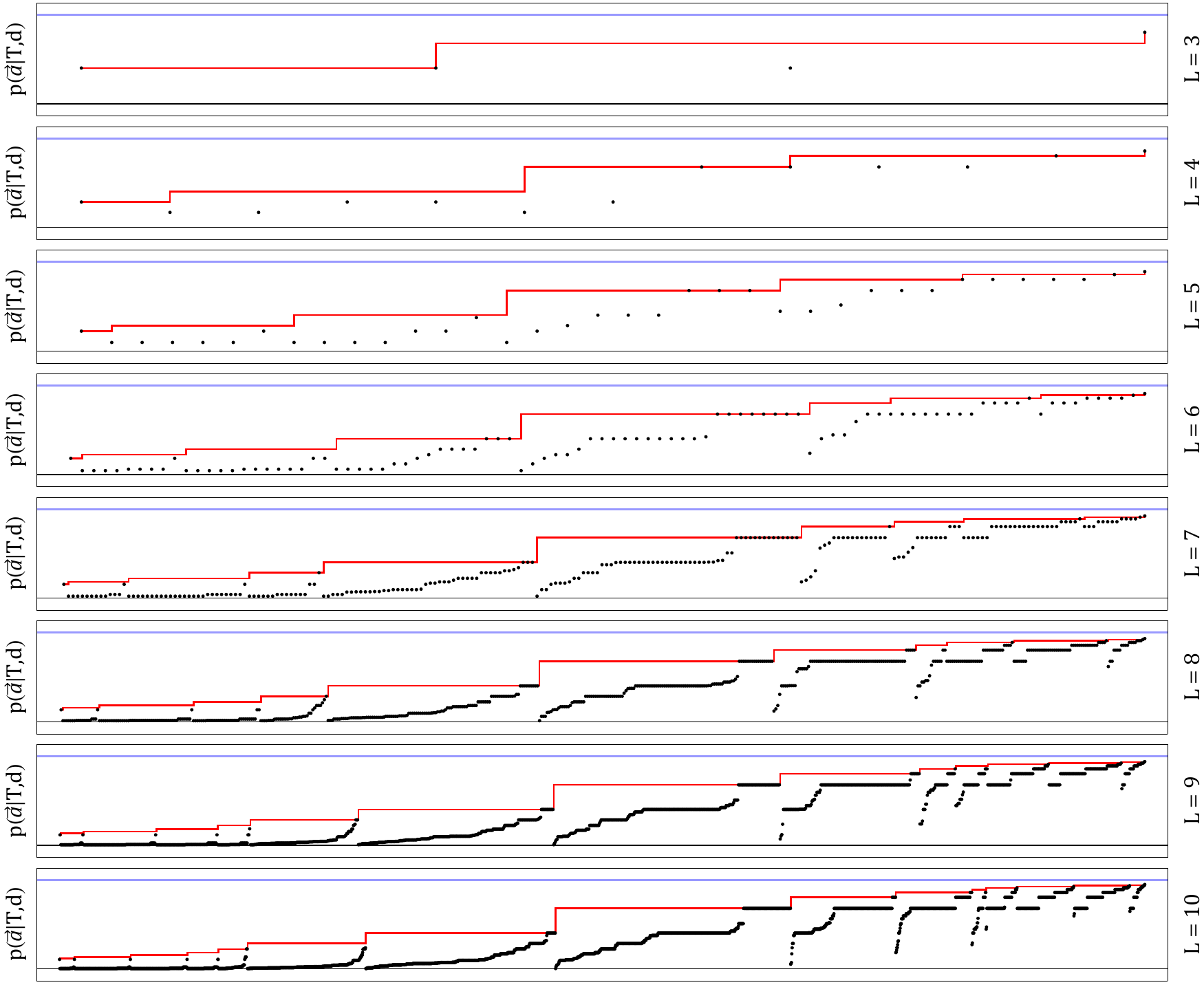}
			\caption{Collected probabilities (black dots) for all sequences $\va$ for models of all dimensions $d < \DC(\va)$ for various lengths $L$. Red line is the upper estimate for different $d$ given by $p(\aot^{\DC(\va)}|E,d)$, based on the enhanced multicyclic models. For readability the data is sorted by increasing probability on the EMCM bound, then by the probability obtained numerically. Blue line is the conjectured $1/\e$ bound. The appearance of ``plateaus'' suggest the existence of additional structures within sequences, giving rise to a few equivalence classes.}
			\label{fig:emcm_upper_bounds}
		\end{figure}
		
		\section{Quantum models}\label{app:quantummodels}
		
		The construction of explicit quantum models is a more difficult task. Recall that in the quantum case transitions are described by the instruments  $\II = (\II_0, \II_1)$ in the Heisenberg picture, i.e., $\II_a$ is completely positive (CP) for $a=0,1$ and $\II_0 + \II_1$ is a unital map. For an initial state $\rho$ on a $d$-dimensional Hilbert space, the probability for a sequence  $\va$ is then $p(\va|\II,d) = \tr[\rho\ \II_{a_1} \circ  \II_{a_2}\circ \ldots \II_{a_L} (\openone) ]$.
		
		We may model this problem numerically by defining the instruments $\II_a$ in terms of their Kraus representation,
		\begin{equation}
			\II_a(\cdot) = \sum_{i=1}^{N_K} \left(K_a^i\right)^\dagger \cdot \left(K_a^i\right),
		\end{equation}
		with $N_K$ the number of Kraus operator for the instrument. For simplicity, we assume the same number of operators for both instruments. These Kraus operators must satisfy the Kraus condition
		\begin{equation}
			\sum_a \sum_{i=1}^{N_K} (K_a^i) (K_a^i)^\dagger = \id.
			\label{eq:krauscond}
		\end{equation}
		This gives us the following constrained optimization problem:
		\begin{equation}\label{eq:optprobquantum}
			\begin{split}
				\max_{\II}\ & p(\va|\II,d) = \tr[\rho\ \II_{a_1} \circ \II_{a_2}\circ \ldots \II_{a_L} (\openone) ].
				\\
				\text{subjected to:} & \sum_{a,i} (K_a^i)^\dagger (K_a^i) = \id.
			\end{split}
		\end{equation}
		To simplify the optimization, we turn this into an unconstrained problem as follows. We first define arbitrary $d \times d$ complex matrices $B_a^i$ as our optimization parameters, then compute the matrix
		\begin{equation}
			E = \sum_{a,i} (B_a^i)^\dagger (B_a^i),
		\end{equation}
		which by construction is positive semidefinite. Let $\lambda_\text{max}$ be the maximum eigenvalue of this matrix. We can now define the (approximate) Kraus operators by the normalization $K_a^i = {B_a^i} / {\sqrt{\lambda_\text{max}}}$. This ensures $E/\lambda_\text{max} \le \id$.
		
		There is no guarantee that these Kraus operators will satisfy the condition in \autoref{eq:krauscond} exactly, but this is not relevant for our optimization purposes, as the optimization procedure will always naturally favor $E / \lambda_\text{max} \approx \id$. This can be interpreted as if our experiment contained a third potential output, $\perp$, whose instrument $\II_\perp$ corresponds to the missing Kraus operators required to complete the Kraus condition. Since the output $\perp$ does not occur in the sequences we are considering, the contribution of $\II_\perp$ will vanish in the optimization of probability for the sequences not containing $\perp$.
		
		The objective function in the problem now requires computation of the maximal eigenvalue of $E$ at every evaluation, which is a non-trivial mathematical operation. This required abandoning PyTorch for a custom implementation of the Adam algorithm. We fixed a gradient step size of $10^{-6}$, and for the Adam algorithm, the parameters $\beta_1 = 0.9$, $\beta_2 = 0.999$, $\epsilon = 10^{-8}$, and a learning rate of $0.003$. We performed a numerical survey for all sequences of length $L = 3, \dots, 7$, with $d = 1, \dots, \DC(\va) - 1$ and $N_K = 1, 2, 3$. For every scenario, we computed $30$ random trials for a fixed $5000$ iterations, and saved the best models and probabilities found in each set of trials. Some results are shown in \autoref{fig:quantumperf}, where we omit sequences with trivial models ($d = 1$).
		
		\begin{figure}\centering
				\begin{minipage}[b]{0.45\linewidth}
						\includegraphics[width=\linewidth]{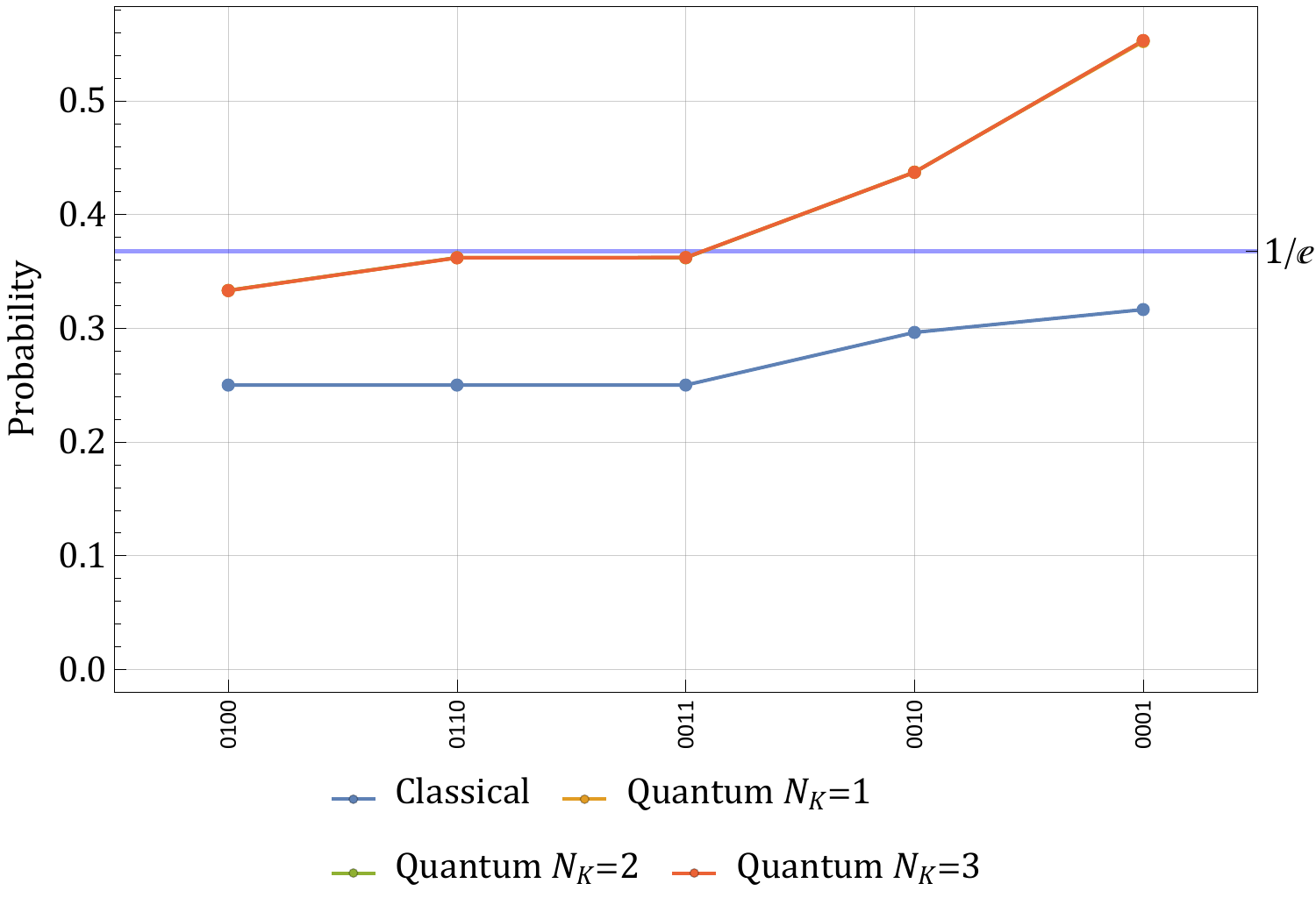}
					\end{minipage} \quad
				\begin{minipage}[b]{0.45\linewidth}
						\includegraphics[width=\linewidth]{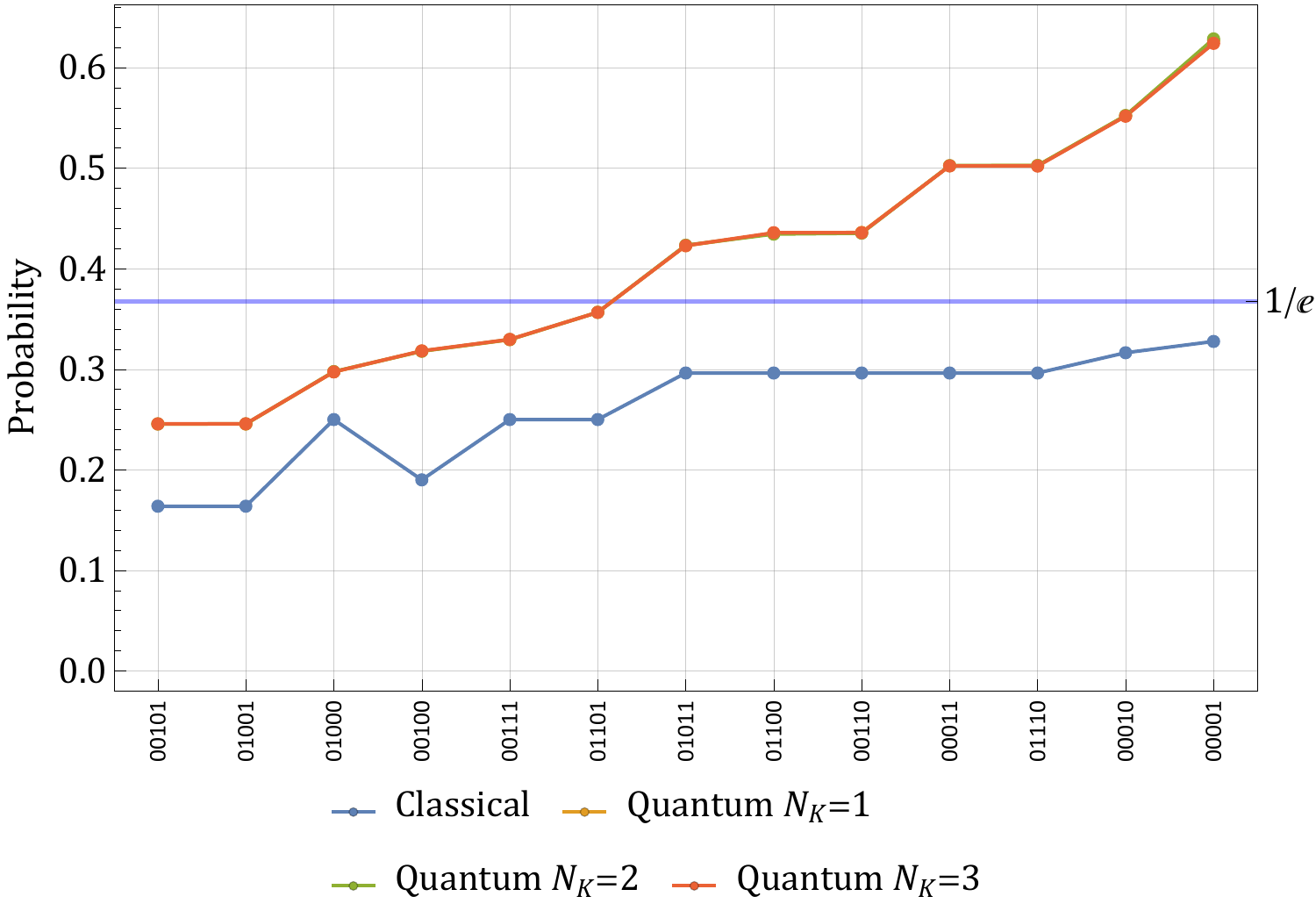}
					\end{minipage} \\
				\begin{minipage}[b]{0.45\linewidth}
						\includegraphics[width=\linewidth]{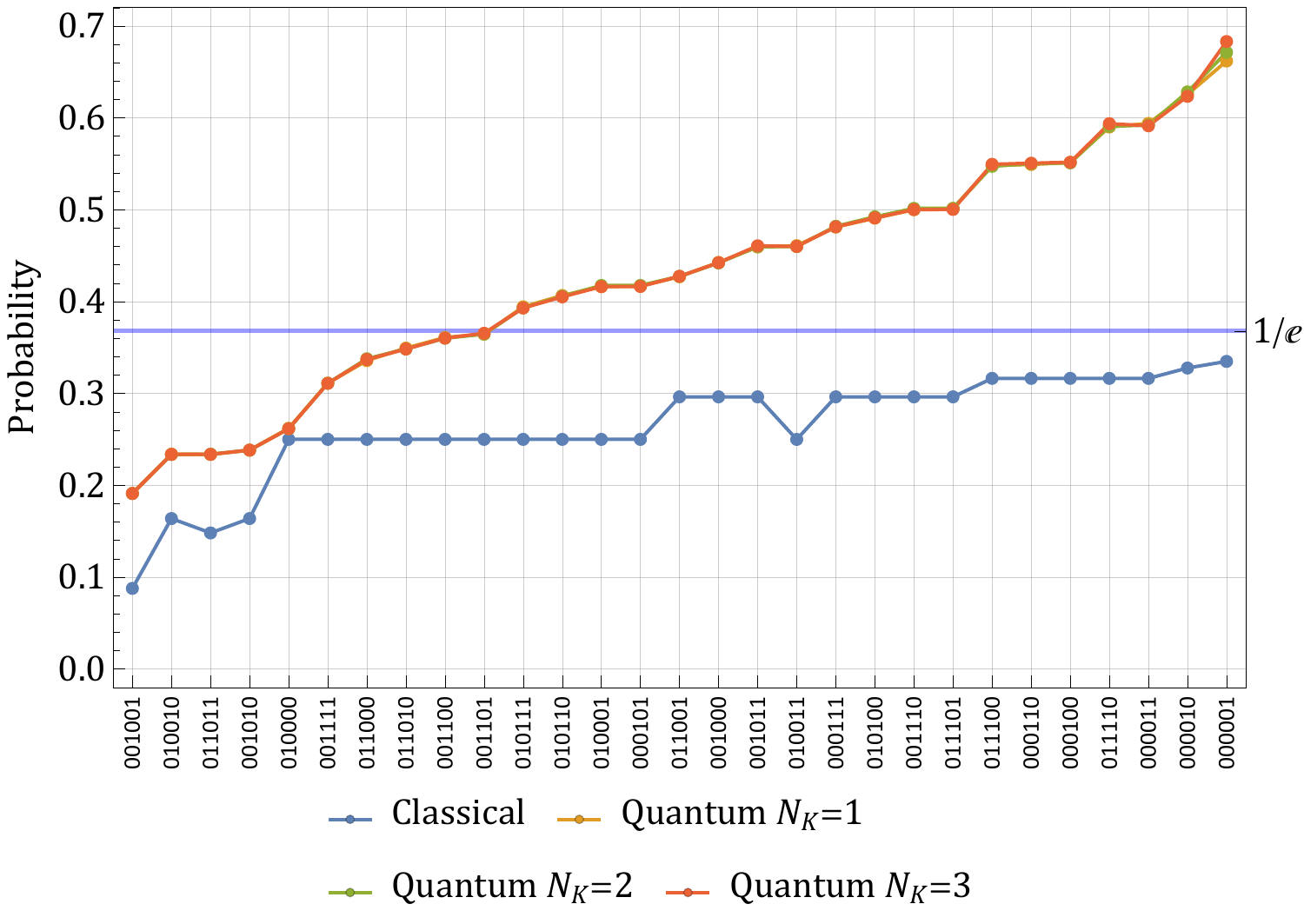}
					\end{minipage} \quad
				\begin{minipage}[b]{0.45\linewidth}
						\includegraphics[width=\linewidth]{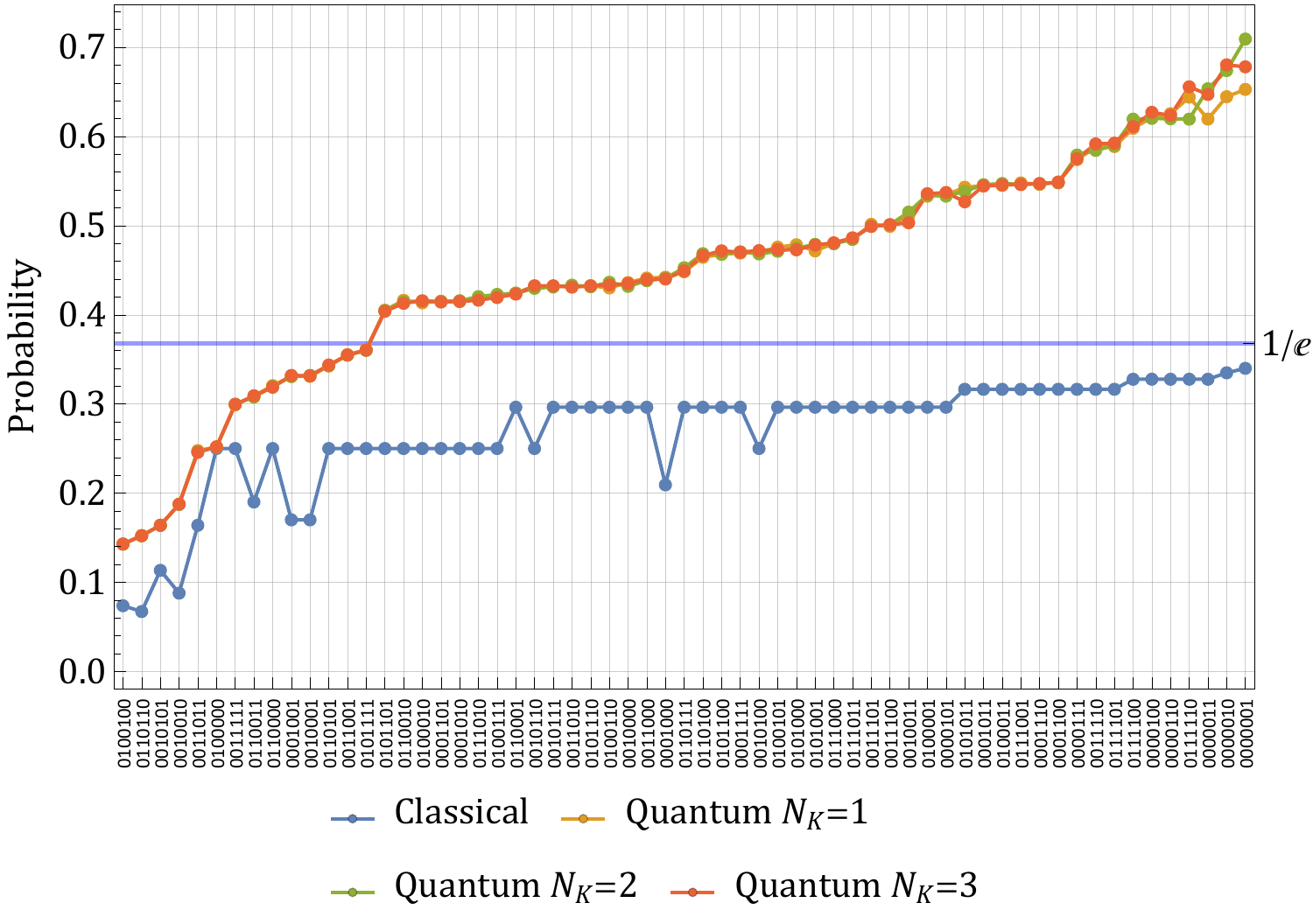}
					\end{minipage}
				\caption{Performance of quantum models for sequences (up to the $0 \leftrightarrow 1$ symmetry) of lengths $L = 4, \dots, 7$, with $d = \DC(\va) - 1 \ge 2$. Sequences were sorted by their quantum performance, and sequences with $\DC(\va) < 3$ were omitted.}
				\label{fig:quantumperf}
		\end{figure}
	
		\begin{figure}\centering
			\includegraphics[width=0.65\linewidth]{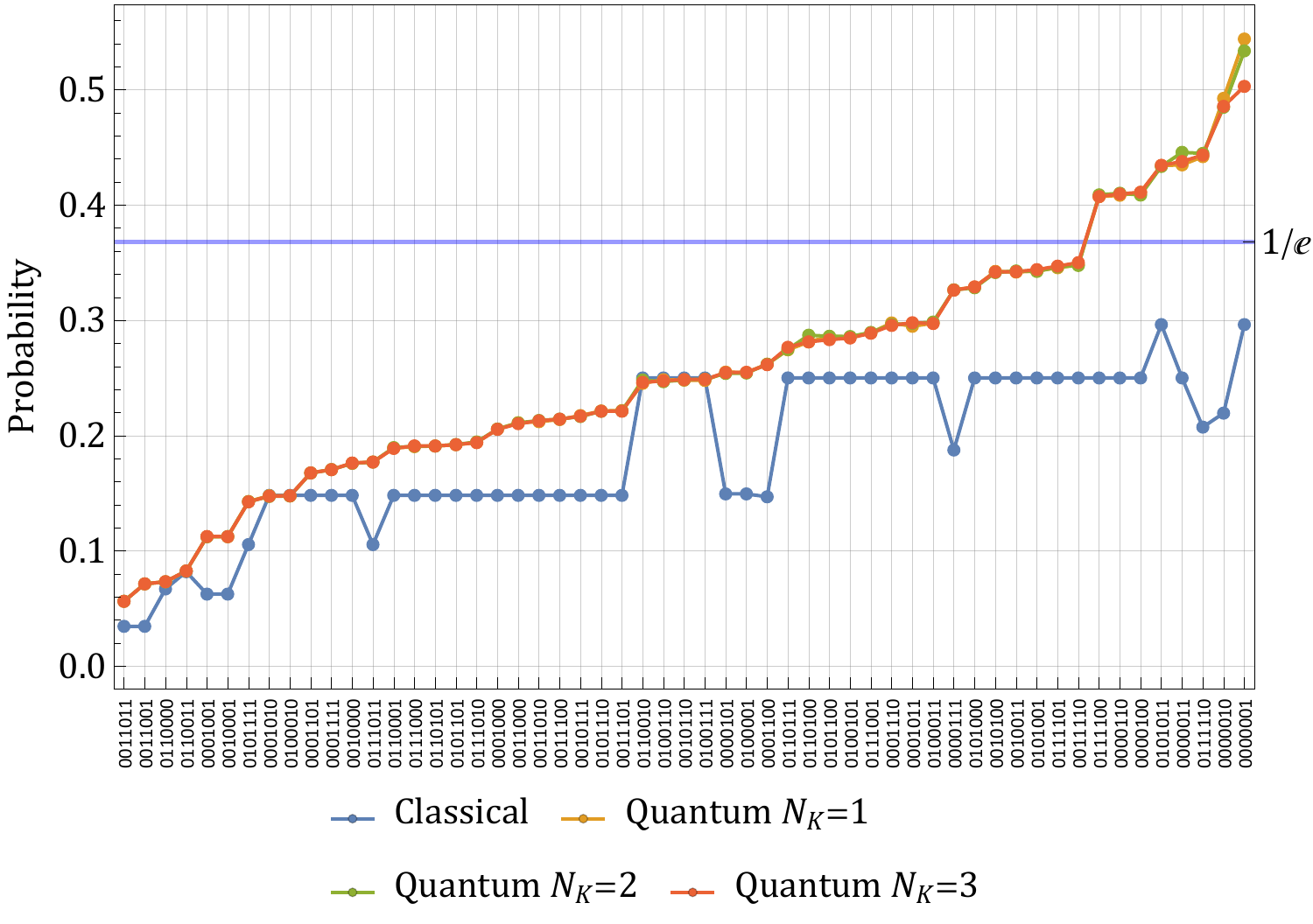}
			\caption{As above, but for $L = 7$ and $d = \DC(\va) - 2 \ge 2$, and sequences with $\DC(\va) < 4$ omitted. Despite the limited number of states, the quantum models still outperform their classical counterparts for most sequences.}
			\label{fig:quantumperf2}
		\end{figure}
	
		As seen in \autoref{fig:quantumperf2}, even with two states missing from a sequence's Deterministic Complexity, many sequences still manage to exceed the classical $1/e$ bound. Additionally, despite the inherent differences in behavior from quantum and classical models, many sequences are shown to have similar performances in both cases, perhaps indicating scenarios in which quantum advantages do not exist. Further investigation is required to understand this behavior.
		
		\subsection{No nontrivial universal quantum bound}
		
		In the following, we discuss the optimization of the probability for the $\aot^L$ sequence for the special case $L=d+1$, i.e., ${d=\DC(\aot^L)-1}$ and the model described above. This optimization was performed in search for an analogous universal quantum bound as suggested in the classical case.
		
		To make the discussion self-contained, let us recall some equations from the main text. For $d=L-1$ and an initial state $\rho=\ketbra{0}$, the probability can be written as
		\begin{align}\label{eq:potq}
		p(\aot^L) &= \tr\left[ K_0^d \ketbra{0} (K_0^\dagger)^d E_1\right] = \tr\left[  \ketbrac{0} (K_0^\dagger)^d E_1 K_0^d \right] \notag \\ &= (1-q) \tr\left[  \ketbra{0} (K_0^\dagger)^d \ketbra{d-1} K_0^d \right] = (1-q) \left| \mean{d-1| K_0^d |0}\right|^2.
		\end{align}
		
		For a fixed $d$, the probability as a function of $\theta_0$ is characterized by a series of $d$ equally-spaced peaks, which become sharper with increasing $d$ (\autoref{fig:qowthetalandscape}), indicating that the optimal value requires fine-tuning.
		
		\begin{figure}[H]\centering
			\includegraphics[width=1\linewidth]{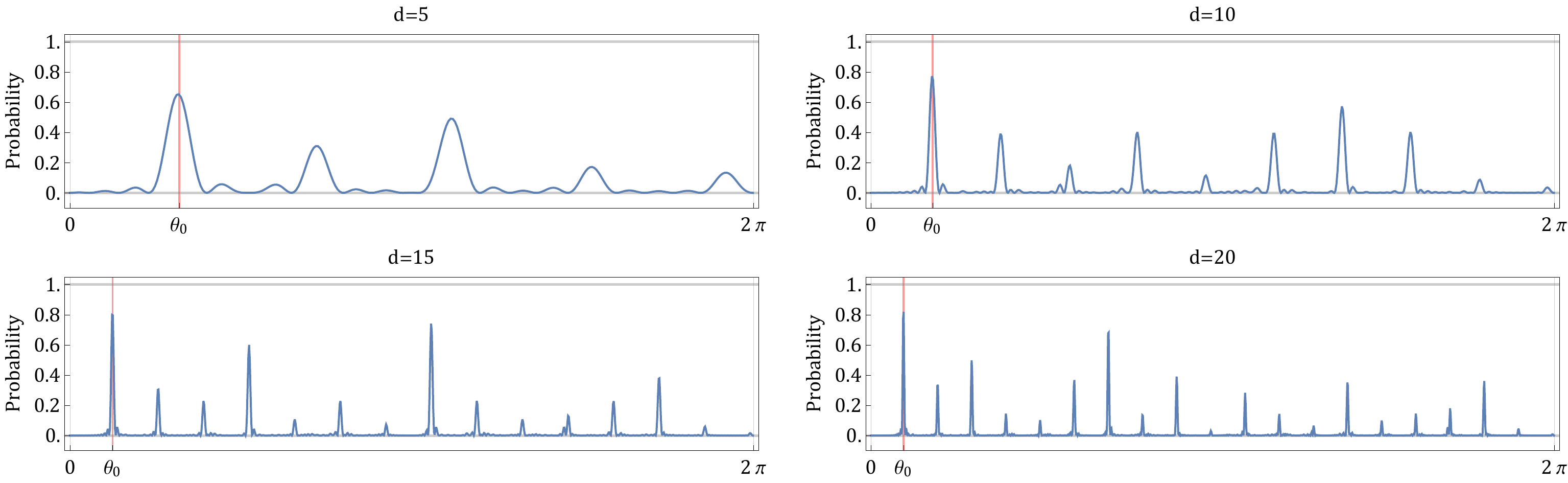}
			\caption{The probability $p(\theta_0|d)$ of the quantum one-way model for increasing $d$. The various peaks become increasingly sharp for higher dimensions. The value $\theta_0 = (2\pi/d)(1-1/d)$ is shown, which appears to be asymptotically optimal.}
			\label{fig:qowthetalandscape}
		\end{figure}
		
		The location of the first peak, which is always optimal, can be estimated by ignoring the effect $E_0$, i.e., $E_0 = \id$ and simply using $K_0 = U_0$. In this case, we obtain
		\begin{equation}
		U_0^d = \e^{-\mathrm{i} H d \theta_0} = F \e^{-\mathrm{i} D d \theta_0} F^\dagger, \quad \text{with}\; D = \operatorname{diag}(0,1,\dots,d-1).
		\end{equation}
		The action of the unitary $U_0^d$ can, then, be computed as
		\begin{equation}
		\left| \bra{d-1}U_0^d\ket{0} \right|^2 = \left| \sum_{k=0}^{d-1} [F]_{d-1,k} \, \e^{-i k d \theta_0} \, [F^\dagger]_{k,0} \right|^2 = \frac{1}{d^2} \left| \sum_{k=0}^{d-1} \exp\left[\frac{2 \pi i}{d} (d-1) k - i k d \theta_0 \right] \right|^2.
		\end{equation}
		As stated in the main text, the maximum occurs when all the amplitudes are in phase, which implies the slowest rotating amplitude ($k = 1$) has to reach zero phase, i.e., when $\frac{2 \pi}{d} (d-1) - d \theta_0 = 0$, giving $\theta_0 = (2\pi/d)(1-1/d)$, as desired.
		
		This simplification made the problem tractable, namely, we were able to compute the probability for high dimension, without the need of an optimization over the parameters $\theta_0$ and $q$. This result, however,  slightly  deviates from the true optimal $\theta_0$ for small $d$. For large $d$, it is asymptotically close to the true optimal. The optimal probability tends towards $1$ as the dimension (and the length of the sequence) increases (\autoref{fig:quantum-ot}), suggesting that there is no nontrivial quantum upper bound, strictly smaller than $1$, in the quantum case.
		
		\subsection{Comparison between classical and quantum one-way models}
		
		\begin{figure}[H]\centering
			\includegraphics[width=0.6\linewidth]{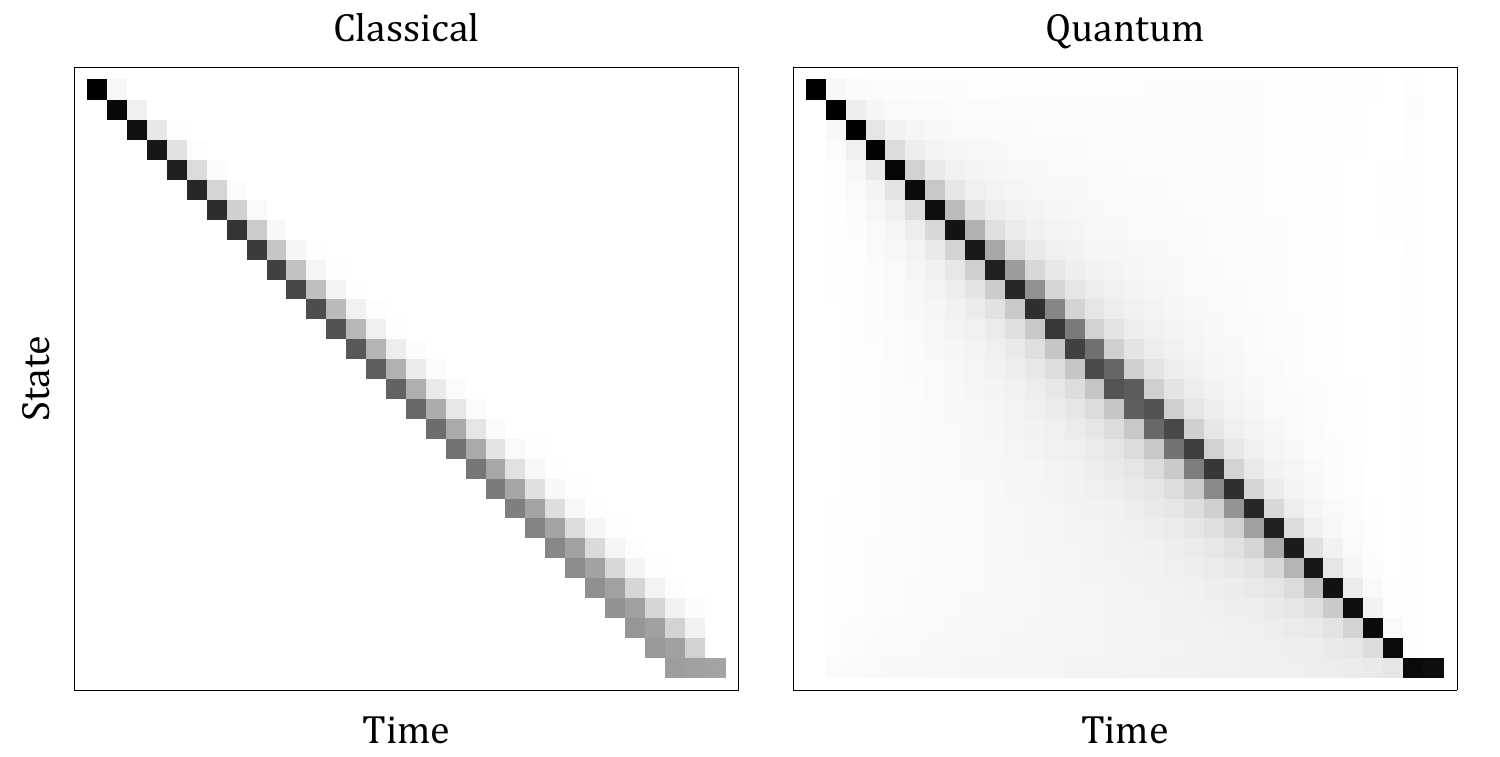}
			\caption{The overall behavior of classical and quantum one-way models, showing the evolution of the probability distributions over the $d$ states. In this diagram, each column represents the states at a given time step in the sequence, starting from the left with $a_1 = 0$ and ending at the right with $a_L = 1$. The first state is at the top. In the quantum case, the distribution corresponds to the diagonal entries in its density matrix. White corresponds to zero probability, black to $1$.}
			\label{fig:owstateevolution}
		\end{figure}
		
		The above results suggest an interesting comparison and intuition between the classical and quantum behaviors in these one-way models.
		
		In the classical case, each transition produces a statistical mixture of states, in such a way that the average behavior is the machine transitioning forward $d$ states after $d+1$ transitions, introducing a delay of one state at the end of the sequence. These transitions always result in a spreading of the probability over many states.
		In the quantum case, the dynamics is governed by the unitary, which produces a coherent superposition of states. The machine is still transitioning forward $d$ states after $d+1$ transitions, but the coherence allows the probability to concentrate at the final state after an initial spread.
		
		The two behaviors are shown in \autoref{fig:owstateevolution}, where the ``concentration'' of probability can be observed in the quantum case.
		
	\end{widetext}

\bibliography{biblio}{}

\end{document}